\documentclass[lettersize,journal]{IEEEtran}
\IEEEoverridecommandlockouts

\usepackage[utf8]{inputenc} 
\usepackage[T1]{fontenc}
\usepackage{url}
\usepackage{ifthen}
\usepackage{cite}
\usepackage[cmex10]{amsmath} 
\usepackage{graphicx}
\usepackage{amsmath,amssymb,amsthm}
\usepackage{cite,soul}
\usepackage{tikz}
\usepackage[capitalize]{cleveref}
\usepackage{diagbox}
\usepackage{multirow}
\usepackage{rotating}
\usepackage{threeparttable}
\usepackage{comment}
\usepackage{enumitem}
\usepackage{algorithm}
\usepackage{algpseudocode}
\usepackage{booktabs}
\DeclareMathOperator{\expec}{\mathbb{E}}

\def\BibTeX{{\rm B\kern-.05em{\sc i\kern-.025em b}\kern-.08em
    T\kern-.1667em\lower.7ex\hbox{E}\kern-.125emX}}

\newcommand{\prob}[1]{{\mathbb P}}
\begin{document}
\title{ Uncertain Location Transmitter and UAV-Aided Warden Based LEO Satellite Covert Communication Systems}



\author{Pei Peng,~\IEEEmembership{Member,~IEEE},   Xianfu Chen,~\IEEEmembership{Senior~Member,~IEEE}, Tianheng  Xu,~\IEEEmembership{Member,~IEEE}\\  Celimuge Wu, ~\IEEEmembership{Senior~Member,~IEEE}, Yulong Zou,~\IEEEmembership{Senior~Member,~IEEE}, Qiang Ni, ~\IEEEmembership{Senior~Member,~IEEE}\\ and Emina Soljanin,~\IEEEmembership{Fellow,~IEEE}

\thanks{Manuscript received \today. \emph{(Corresponding author: Xianfu Chen)}}
\thanks{Pei Peng and Yulong Zou are with the School of Telecommunication and Information Engineering, Nanjing University of Posts and Telecommunications, Nanjing, Jiangsu 210003, China (Emails: pei.peng@njupt.edu.cn; yulong.zou@njupt.edu.cn).}%
\thanks{Xianfu Chen is with Shenzhen CyberAray Network Technology Company Ltd., Shenzhen Guangdong 518042, China (Email: xianfu.chen@ieee.org).}
\thanks{Tianheng Xu is with the Shanghai Advanced Research Institute, Chinese Academy of Sciences, Shanghai 201210, China (Email: xuth@sari.ac.cn).}
\thanks{Celimuge Wu is with the Meta-Networking Research Center, The University
of Electro-Communications, Tokyo 182-8585, Japan (E-mail: celimuge@uec.ac.jp).}
\thanks{Qiang Ni is with the School of Computing and Communications, Lancaster University, LA1 4WA
Lancaster, U.K. (Email: q.ni@lancaster.ac.uk)}
\thanks{Emina Soljanin is with the Department of Electrical and Computer Engineering, Rutgers, The State University of New Jersey, Piscataway, NJ 08854, USA (Email: emina.soljanin@rutgers.edu)}

}%

\markboth{IEEE JOURNAL on Selected Area in Communications, NO. XX, MONTH YY, 25}%
{Shell \MakeLowercase{\textit{et al.}}: A Sample Article Using IEEEtran.cls for IEEE Journals}

\maketitle

\begin{abstract}

We propose a novel covert communication system in which a ground user, Alice, transmits unauthorized message fragments to  Bob, a low-Earth orbit satellite (LEO), and an unmanned aerial vehicle (UAV) warden (Willie) attempts to detect these transmissions.  The key contribution is modeling a scenario where Alice and Willie are unaware of each other's exact locations and move randomly within a specific area. Alice utilizes environmental obstructions to avoid detection and only transmits when the satellite is directly overhead. LEO satellite technology allows users to avoid transmitting messages near a base station.  We introduce two key performance metrics: catch probability (Willie detects and locates Alice during a message chunk transmission) and overall catch probability over multiple message chunks. We analyze how two parameters impact these metrics: 1) the size of the detection window and 2) the number of message chunks. The paper proposes two algorithms to optimize these parameters. The simulation results show that the algorithms effectively reduce the detection risks. This work advances the understanding of covert communication under mobility and uncertainty in satellite-aided systems.

\end{abstract}


\section{Introduction}

\IEEEPARstart{W}{ith} wireless communication technologies being widely used in daily life, information security is essential to protect users' anonymity and privacy\cite{alsabah20216g}. Generally, not only the content of information but also the transmission behavior is valuable to a malicious user\cite{he2017covert}. Covert communication aims to hide the transmission behavior and make wireless transmissions undetectable\cite{chen2023covert}.
In covert communications, the transmitter, Alice, transmits to the receiver, Bob, and the warden, Willie, monitors and analyzes the received signals to detect whether the transmission is happening (see e.g., \cite{peng2022covert}). An information-theoretic approach to achieve covertness by concealing messages as noise is proposed in \cite{bash2013limits, bash2015hiding}. Afterwards, many scenarios focus on extending the model. For example, a jammer can be introduced to this model to guarantee that the received signal power varies randomly \cite{du2022performance,he2023channel,chen2024achieving}; a third participant, Carol, can provide the signal for Alice to hide in it \cite{yu2024covert,wang2024star,li2024covert}. 
Other schemes are considered in a recent comprehensive survey \cite{chen2023covert} and references therein.

Non-Terrestrial Networks (NTNs) are an essential technology of the sixth-generation (6G) communication infrastructure, providing ground users with ubiquitous connectivity without being limited by terrain and topography \cite{feng2023radio,you2024ubiquitous}. 
Low-Earth-Orbit (LEO) satellites are the crucial components of NTNs\cite{wang2024sustainable,capez2024use}. In recent years,  thousands of LEO satellites have been placed around the Earth at altitudes of about 500 to 2000 km above sea level at high speeds to provide global users with relatively low-latency high-bandwidth communications \cite{wang2023satellite,ma2021uav,peng2024blocked}. The deployment of NTNs allows the ground user to use small devices, such as smartphones, to transmit messages to the LEO satellite\cite{he2024direct}. It presents a significant challenge to security \cite{feng2024covert,jia2025robust}. For example, a spy can covertly leak an unauthorized message to the outside through the LEO satellite, considering the vast coverage provided by the satellite. In contrast, we can also use LEO satellite communication to hide various aspects of communications and increase anonymity and privacy.

This paper considers the transmission of covert messages in an LEO satellite communication scenario. The ground user, who acts as Alice, wants to transmit an unauthorized message to the LEO satellite, which acts as Bob. An unmanned aerial vehicle (UAV) acts as Willie, guarding the area to detect unauthorized transmissions \cite{chen2021uav}. We assume that Alice and Willie realize each other's existence but do not know each other's location \cite{zhou2019joint}. Alice will be static during the transmission, but she will move and randomly find another location for the subsequent transmission. Since satellite communication is allowed in this area, Willie has to move close to Alice to judge whether the transmission is authorized, even though he detected the transmission. 

Various covert satellite communication scenarios were recently considered in \cite{feng2024covert,jia2025robust,yu2024covert,song2023ris,liu2024covert}. For example, \cite{feng2024covert} achieves the covert transmission from the satellite to the user on the ground by deploying massive LEO satellites at different altitudes around the Earth as a backhaul for UAVs that serve users. 
\cite{jia2025robust} achieves the covert transmission by employing a rate-splitting multiple access technique at the satellite transmitting to users and simultaneously transmitting jamming signals to Willie. \cite{yu2024covert} achieves the covert transmission by letting the satellite send an overt signal superimposed by a covert signal, which the public user Carol, the receiver Bob, and the warden Willie receive. \cite{song2023ris} achieves the covert transmission with the help of the reconfigurable intelligent surface (RIS), which is placed near the receiver Bob to reflect the received signal.
A growing body of literature discusses covert communication in UAV scenarios, e.g., \cite{chen2021uav,du2022performance,wang2022covert,wang2024star,jiao2024uav,zhou2021three}, and references therein. In this case, \cite{chen2021uav} used a UAV as a relay to help Alice transmit the message, and Alice used the obstructions to weaken Willie's detection. \cite{du2022performance} considered a jammer-aided UAV covert communication system where the jammer helps the UAV Alice to transmit covertly by interfering with Willie. \cite{wang2024star} considered a UAV as Alice, and Carol combined simultaneously transmitting and RIS to help Alice cover the message.
Some papers also consider the movements of Alice, Bob, and Willie. For example, \cite{zhou2019joint} considered the UAV scenario in which both Bob and Willie are UAVs and neither knows the location of the other. \cite{zheng2019multi} considered multiple Willies move randomly and independently inside a certain region around Bob in the ground scenario. \cite{jiang2021resource,jiang2021covert,deng2024joint} considered the UAV scenario, and  Alice performs as a UAV, Bob and Willie are static on the ground.

Most relevant papers do not analyze how location uncertainty affects the detection probability or only assume that the moving character is the UAV. In contrast, the characters on the ground are static. This is because the transmission channel can vary rapidly with different environments and be significantly affected by surrounding obstructions on the ground. 

We consider the impact of location uncertainty on detection probability by assuming that Alice and Willie's locations change randomly in the area.
Here, we focus on how Willie's detection probability changes with different channel coefficients and how Alice's transmission rate affects Willie's judgment of unauthorized transmission after he detects a transmission. 
The contributions of this paper are summarized as follows:
\begin{enumerate}[leftmargin=*]
\item We propose a novel covert communication scenario that considers using the satellite to transmit and the UAV to detect. In this model, we relate the transmission rate, the detection probability, and the UAV's moving speed; all the parameters affect the covert transmission performance. 

\item We introduce the catching probability as the new performance metric, defined as the probability that Willie detects the transmission and locates Alice during the transmission, to evaluate the covertness of the transmission, which is based on the assumption that Alice and Willie realize each other's existence but do not know each other's location. 

\item We derive the expression of the catch probability, analyze it, and propose approximate algorithms to find the optimal detection window size accordingly. The simulation results show that the catching probability changes with the detection window size and that the proposed algorithm can accurately approximate the optimal value.

\item We analyze how many chunks should be split to minimize the overall probability of catching and propose an algorithm to approximate the number of chunks that should be split. The simulation results show that the number of chunks split significantly affects the overall probability of catching, and the proposed algorithm performs well.
\end{enumerate}

The remainder of the paper is organized as follows. In Section~\ref{sec:sys}, we describe the architecture of the covert communication scenario and formulate the problem. In Section~\ref{sec:willie}, we derive and analyze the expression of the catching probability and propose an algorithm to approximate the optimal detection window size.
In Section~\ref{sec:alice}, we derive and analyze the expression of the overall probability of catching and the message splitting methods in two cases.
Numerical and simulation results are given in Section~\ref{sec:simulation}.
Finally, our work is concluded in Section~\ref{sec:conclusions}.

\section{ System Model}
\label{sec:sys}

\begin{figure}[htb]
    \centering
    \includegraphics[width=0.5\textwidth]{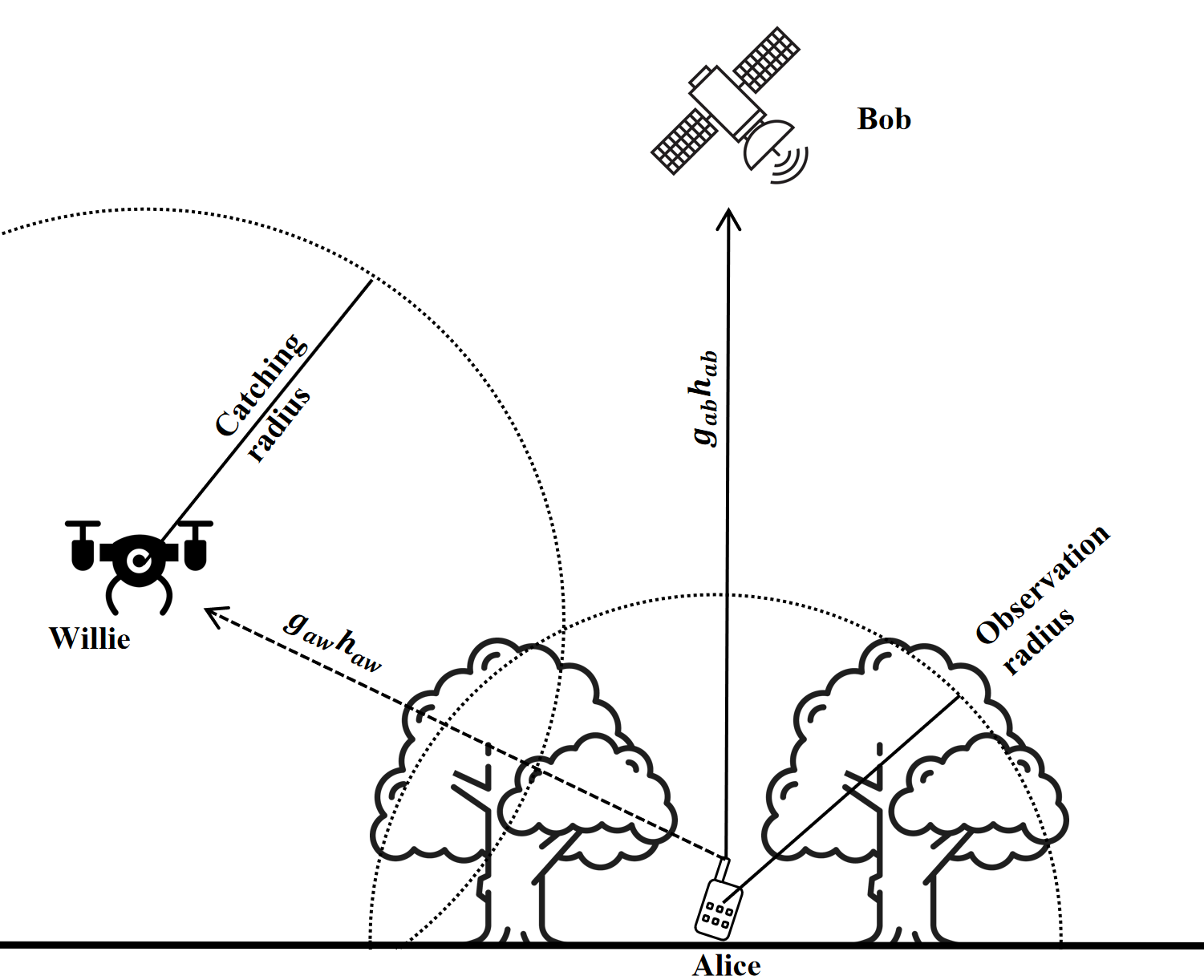}
    \caption{Covert communication model in the LEO satellite communication systems.  }
    \label{fig:systemmodel}
\end{figure}

 As shown in Fig~\ref{fig:systemmodel}, Alice moves on the ground and wants to covertly transmit an unauthorized message to Bob, an LEO satellite. Willlie, the warden, is a UAV patrolling the sky and trying to detect potential transmissions. Alice and Bob know each other's existence but do not know each other's location. Since satellite communication is not forbidden in this area, Bob has to move close to Alice during her transmission to judge whether the transmission is authorized. We assume that Alice and Willie move in a square area of length $u$ \cite{deng2024joint}. Obstructions, such as buildings and trees, are commonly located in this area. 

\subsubsection{Alice} 
Alice can take advantage of the obstructions in the area, such as near a tree or a building, to reduce Willie's detection probability \cite{chen2021uav}. Before transmitting, she will observe the surroundings. If Willie appears in the radius $r_a$, she will stop transmitting to avoid Willie's detection.
Of course, Alice's transmission may also be affected by the obstructions. Thus, she will only transmit the message when the LEO satellite is in the central sky; Alice has limited transmission time in a transmission slot. After the previous transmission, she moves randomly and finds another appropriate location to transmit in the next transmission slot.

 \subsubsection{Bob} 
Bob is the LEO satellite that receives Alice's message. However, the LEO satellite communication system is organized by multiple satellites, and Bob does not represent a specific satellite. 
 The LEO satellite is moving fast around the Earth. Each satellite has a transmission slot of size $t_c$. After time $t_c$, another satellite will move into the sky over Alice. The new satellite will also be Bob. We assume that the LEO satellites are distributed uniformly, then $t_c$ is constant.

 \subsubsection{Willie}
 Willie patrols slowly and detects the transmission periodically only when hovering in the sky. We assume that when he detects the potential transmission, he will fly to the signal source at much higher speeds $v_w$. Willie has a catching radius $r_w$, which is defined as the maximum distance from Alice that Willie can judge the authorization of the transmission. Here, we assume $r_w>r_a$.  
 
\subsection{Channel Model From Alice to Bob}

Although Alice transmits near buildings or trees, obstructions will not affect the transmission since she only transmits when the LEO satellite is overhead. Therefore, we assume that the channel model from Alice to Bob follows the large-scale Line-of-Sight (LoS) path loss $g_{ab}$ with a small-scale Rician fading $h_{ab}$. The channel coefficient between Alice and Bob is as follows.

\begin{equation}
\label{eq:gan_ab}
g_{ab}=\frac{\sqrt{G_a G_b}}{d_{ab}^{\alpha_L}}. 
\end{equation}
Where $G_a$ is the gain of Alice’s transmission antenna, $G_b$ is the gain of Bob's receiving antenna, $d_{ab}$ is the distance between Alice and Bob, and $\alpha_L$ is the path-loss exponent of LoS. Here, we assume that $\alpha_L=1$ reflects the free propagation nature of LEO satellite communication. 

\begin{equation}
h_{ab}=\sqrt{\frac{K_0}{1+K_0}}h_{LoS}+\sqrt{\frac{1}{1+K_0}}h_{NLoS}.
\end{equation}
Where $K_0$ represents the Rician factor, $h_{LoS}$ denotes the gain of the LoS and $||h_{LoS}||=1$, and $h_{NLoS}$ denotes the gain of the LoS and $h_{NLoS}\sim \mathcal{CN}(0,1)$.

Considering the noise at Bob is the complex additive white Gaussian noise (AWGN) $n_b[k]\sim \mathcal{CN}(0,\sigma_b^2)$, the transmission rate between Alice and Bob is 
\begin{equation}
\label{Eq:rate}
    R_{ab}=\log(1+\frac{P_a g_{ab}^2|h_{ab}|^2}{\sigma_{b}^2}).
\end{equation}
Notice that $R_{ab}$ is the transmission rate per Hz. In the following, we use the average transmission rate $\overline{R}_{ab}$ to evaluate Alice's transmission time given the message size. The average transmission rate is calculated by the Monte Carlo simulation. In this paper, we do not consider the specific modulation, and we assume that Alice transmits each symbol as one bit.

\subsection{Channel Model from Alice to Willie and Hypothesis Testing}

Since Alice uses the building or the trees to inhibit Willie from detecting the transmission, we assume the channel model from Alice to Willie follows a large-scale NLoS path loss $g_{aw}$ with a small-scale Rayleigh fading $h_{aw}$. The channel coefficient between Alice and Willie is as follows.

\begin{equation}
\label{eq:gain_aw}
g_{aw}=\frac{\sqrt{\eta G_a G_w}}{d_{aw}^{\alpha_N}}.
\end{equation}
Where $G_a$ is the gain of Alice’s transmission antenna, $G_w$ is the gain of Willie's receiving antenna, $d_{aw}$ is the distance between Alice and Willie, $\alpha_N$ is the path-loss exponent of NLoS, and $\eta$ is the excessive path-loss coefficient of NLoS channels. Here, we assume that $\alpha=1.5$. The coefficient of Rayleigh fading
$h_{aw}\sim \mathcal{CN}(0,1)$.

The baseband signal sent by Alice is denoted as $x_a[k]$, $k=1,2,\dots M$, which satisfies $\expec[|x_a[k]|^2]=1$. Each symbol $x_a[k]$ is independent and distributed identically (i.i.d.) where $M$ is the length of the message. Alice will transmit the signal with a fixed transmission power $P_a$. Then the received signal at Willie is
\begin{equation}
y_{w}[k]=\sqrt{P_a}g_{aw}h_{aw}x_a[k]+n_w[k].
\end{equation}
Here $n_w[k]\sim \mathcal{CN}(0,\sigma_w^2)$ is the noise at Willie, which is also i.i.d. 

We assume that Willie determines whether Alice is transmitting based on the received signal power. Since Willie does not know when Alice starts transmitting, he will detect two kinds of signals. $\mathcal{H}_0$ represents the null hypothesis that Alice is silent. $\mathcal{H}_1$ is the alternative hypothesis that Alice performs the transmission to Bob. Then the received signal at Willie is as follows.
\begin{equation}
\label{eq:receive}
y_w[k] = 
~\begin{cases} 
n_w[k], & \mathcal{H}_0. \\ 
\sqrt{P_a}g_{aw}h_{aw}x_a[k]+n_w[k], & \mathcal{H}_1.
\end{cases}   
\end{equation}

Next, Willie will make a binary decision by comparing the received signal's average power $T_w$ with the preset detection threshold $\Gamma_w$. The formula is given by
\begin{equation}
\label{eq:detect}
    T_w=\frac{1}{L}\sum_{i=1}^L|y_w[i]|^2 \underset{D_0}{\overset{D_1}{\gtrless}} \Gamma_w.
\end{equation}
Where $L$ is the detection window size, representing how many symbols Willie will observe to detect. $D_0$ and $D_1$ denote the detection results. When $T_w>\Gamma_w$, Willie chooses the decision $D_1$, Alice is transmitting to Bob. On the contrary, when  $T_w<\Gamma_w$, Willie chooses the decision $D_0$, that is, Alice is silent.

\subsection{Willie's Catching Process}

\begin{figure}[htb]
    \centering
    \includegraphics[width=0.5\textwidth]{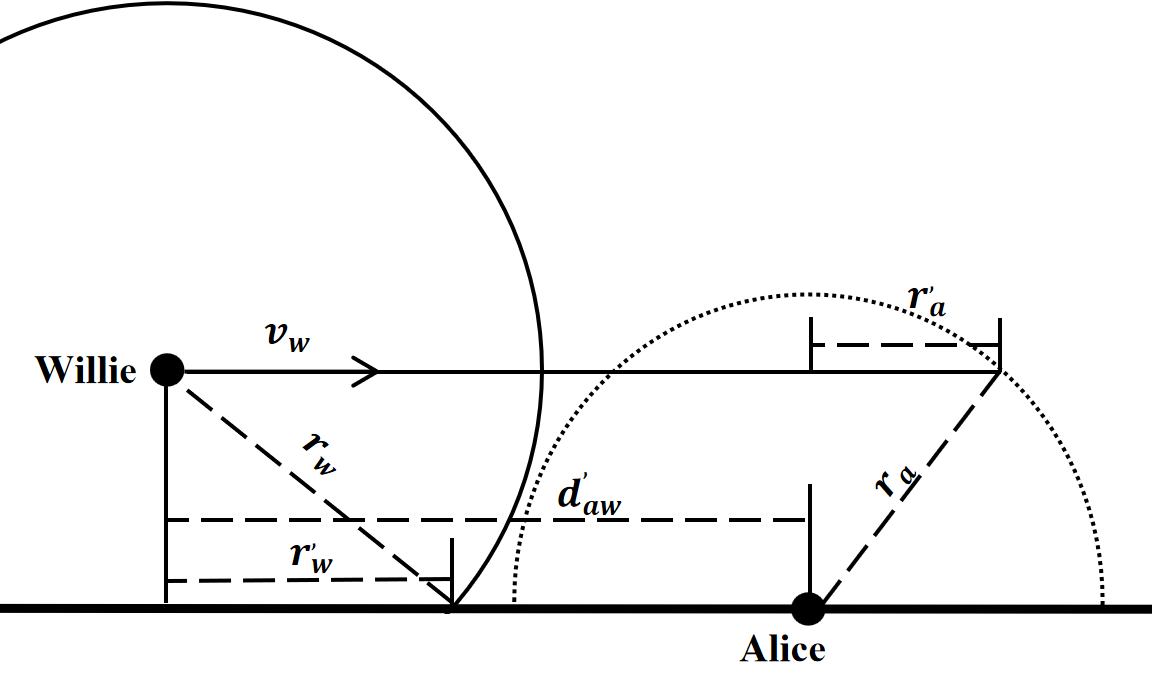}
    \caption{Definitions of Willie's catching radius and Alice's observation radius.  }
    \label{fig:distance}
\end{figure}

As introduced in Fig.~\ref{fig:systemmodel}, Willie must locate Alice during her transmission. Otherwise, he can not tell whether the detected transmission is unauthorized. Fig.~\ref{fig:distance} represents Willie's catching processing after detecting a transmission. In the figure, the parameters are denoted as follows. $v_w$ is the catching speed, $d_{aw}$ is the distance between Alice and Willie, $r_w$ is Willie's catching radius, $r_a$ is Alice's observation radius. $d_{aw}^{\prime}$, $r_{w}^{\prime}$ and $r_a^{\prime}$ are the projections of $d_{aw}$, $r_w$ and $r_a$ to the horizontal plane respectively.
According to Fig.~\ref{fig:distance}, when Willie detects a transmission, he will horizontally fly to the transmission source with a speed $v_w$. He will continuously detect Alice's transmission to locate her during the flight. Once the distance $d_{aw}\le r_w$, Willie will catch Alice if she is still transmitting.

\subsection{Problem Statement and Performance Metrics}
We consider two main problems based on the proposed novel covert communication model. First, how does the size of the detection window affect the system performance in terms of the probability of catching? We define the catching probability $P_{ca}$ as the probability that Willie catches Alice during Alice's transmission. From \eqref{eq:receive} and \eqref{eq:detect}, we know that Willie's detection will be more accurate when the detection window size $L$ is larger. However, Alice is transmitting a finite-length message. If Willie spends too much time receiving the detection symbols, he may not have enough time to locate Alice.

Second, how should Alice split the message chunks to reduce the overall catching probability? Based on the first problem, the catching probability for each chunk can be reduced when Alice splits the message into smaller chunks. However, Alice must transmit multiple times to complete transferring the whole message, which offers Willie more chances to catch Alice. We define the overall catching probability $P_{ov}$ as the catching probability considering multiple times.

\section{Catching Analysis at Willie Considering Detection Window Size}
\label{sec:willie}

In this section, we first analyze the detection probability at Willie and find the minimum miss detection subject to the false alarm constraint. Then, we derive the probability density function and the cumulative distribution function for the distance between Alice and Willie. Finally, we derive the approximated catching probability and propose an algorithm to find the optimal detection window size.

\subsection{Detection Analysis}

According to binary hypothesis testing, Willie may make two types of mistakes. The first is the False Alarm (FA), which represents that Willie makes the decision $D_1$ while Alice keeps silent. The second is the Miss Detection (MD), representing that Willie makes the decision $D_0$ while Alice is transmitting. We will find the expressions of FA and MD in the following.

\subsubsection{False alarm probability $P_{FA}$}
According to \eqref{eq:receive} and \eqref{eq:detect}, when the detection threshold $\Gamma_w$ is given, the false alarm probability $P_{FA}$ can be expressed as
\begin{equation}
\label{eq:false}
    P_{FA}=Pr(T_w\ge\Gamma_w|\mathcal{H}_0).
\end{equation}

When Alice is at status $H_0$, the transmission does not happen; Willie will only receive the complex AWGN $n_w[k]\sim \mathcal{CN}(0,\sigma_w^2)$. Considering the complex AWGN, we have $2|n_w[k]|^2/\sigma_w^2$, which follows the chi-square distribution. Then we have
\begin{equation}
    \frac{|y_w[k]|^2}{\sigma_w^2/2}\sim \chi^2(2).
\end{equation}
Where $\chi(\cdot)$ is the chi-square distribution. Since noise symbols $n_w[k]$ ($k=1,2,\dots M$) are i.i.d., the average Power $T_w$ can be expressed as

\begin{equation}
\label{eq:avpower}
    T_w=\frac{1}{L}\sum_{i=1}^L|y_w[i]|^2 \sim \frac{\sigma_w^2}{2L}\chi^2(2L).
\end{equation}

Then, the false alarm probability $P_{FA}$ can be calculated as follows.

\begin{equation}
\begin{aligned}
\label{eq:FA}
     P_{FA}&=\int_{\Gamma_w}^{\infty}\frac{T_w^{L-1}}{\Gamma(L)}(\frac{L}{\sigma_w^2})^Le^{-\frac{LT_w}{\sigma_w^2}}d T_w\\
     &=1-\frac{\gamma(L,\frac{L\Gamma_w}{\sigma_w^2})}{\Gamma(L)}.
\end{aligned}
\end{equation}
Where $\Gamma(L)=(L-1)!$ is the Gamma function. $\gamma(\cdot,\cdot)$ is the lower incomplete Gamma function given by $\gamma(n,x)=\int_{0}^{x}e^{-z}z^{n-1}dz$.

\subsubsection{Miss detection probability $P_{MD}$ }

When the detection threshold $\Gamma_w$ is given, the miss detection probability $P_{MD}$ can be expressed as
\begin{equation}
    P_{MD}=Pr(T_w<\Gamma_w|\mathcal{H}_1).
\end{equation}

Considering that Willie is assumed to hover during the detection, the coefficient of Rayleigh fading $h_{aw}\sim \mathcal{CN}(0,1)$ does not vary over time. In addition, since the signal symbols $x_a[k]$ and the complex AWGN $n_w[k]$ ($k=1,2,\dots M$) are i.i.d., the received signal $y_w[k]$ also follows the complex Gaussian distribution. That is, the expression of $y_w[k]$ is given by

\begin{equation}
    \sqrt{P_a}g_{aw}h_{aw}x_a[k]+n_w[k]\sim \mathcal{CN}(0,g_{aw}^2P_a+\sigma_w^2).
\end{equation}

Then the average power of the received signal $T_w$ at the status $\mathcal{H}_1$ is
\[
T_w^{\mathcal{H}_1} = \frac{1}{L}\sum_{i=1}^L|\sqrt{P_a}g_{aw}h_{aw}x_a[k]+n_w[k]|^2.
\]

Similar to \eqref{eq:avpower}, the average power $T_w$ also follows the chi-square distribution. Then, we can get its expression as
\begin{equation}
    T_w^{\mathcal{H}_1} \sim \frac{g_{aw}^2P_a+\sigma_w^2}{2L}\chi^2(2L).
\end{equation}

Then, the probability of missdetection $ P_{MD}$ can be calculated as follows.

\begin{equation}
\begin{aligned}
\label{eq:MD}
     P_{MD}&=\int_{0}^{\Gamma_w}\frac{T_w^{L-1}}{\Gamma(L)}(\frac{L}{g_{aw}^2P_a+\sigma_w^2})^Le^{-\frac{LT_w}{g_{aw}^2P_a+\sigma_w^2}}d T_w\\
     &=\frac{\gamma(L,\frac{L\Gamma_w}{g_{aw}^2P_a+\sigma_w^2})}{\Gamma(L)}.
\end{aligned}
\end{equation}

\subsubsection{Detection threshold $\Gamma_w$} 
Since Willie's catching probability is mainly affected by the MD probability, we mainly focus on minimizing $P_{MD}$ with the threshold $L$. However, when the FA probability is high, Willie will waste more energy chasing the wrong target and may also miss the chance to detect Alice's transmission.  Therefore, $P_{FA}$ should be constrained by a minimal value $\delta$. Then we need to solve the optimization problem $\min_{\Gamma_w} P_{MD}$ subject to $ P_{FA}\le \delta$. 

According to \eqref{eq:FA} and \eqref{eq:MD}, we find that $P_{FA}$ decreases with the increasing $\Gamma_W$ and $P_{MD}$ increases with $\Gamma_W$. Since the FA probability is constrained by $ P_{FA}\le \delta$, the optimal threshold $L_w^*$ can be calculated by
\begin{equation}
    \frac{\gamma(L,\frac{L\Gamma_w}{\sigma_w^2})}{\Gamma(L)}= 1-\delta.
\end{equation}

According to the definitions of FA and MD, $P_{FA}$ and $P_{MD}$ decrease with the increasing detection window size $L$. Then the optimal threshold $L_w^*$ decreases with the increasing $L$.

\subsection{Distance Analysis Between Alice and Willie}
According to the system model, Alice and Willie move in a square area of length $u$. During the transmission slot, Alice is static in a location. Bob also remains static during the detection. Other than that, they will move randomly in the square area. According to \eqref{eq:gain_aw}, the coefficient $g_{aw}(d_{aw})$ can be considered as a function of $d_{aw}$, where $d_{aw}$ is the distance between Alice and Willie. Therefore, it is essential to find its expression.
Let Alice's location and Willie's projected location in the horizontal plane be $(u_{ax},u_{ay})$ and $(u_{wx},u_{wy})$, respectively. Then $u_{ax}$, $u_{ay}$, $u_{wx}$ and $u_{wy}$ follow the uniform distribution $U(0,u)$. The absolute values of $d_x=|u_{ax}-u_{wx}|$ and $d_y=|u_{ay}-u_{wy}|$ follow a triangle distribution with the probability density function (PDF):
\begin{equation}
    f(x)=\frac{2(u-x)}{u^2}, 0\le x\le u.
\end{equation}
Then the joint PDF is 
\begin{equation}
    \label{eq:trian}
    f(d_x,d_y)=\frac{4(u-d_x)(u-d_y)}{u^4}, 0\le d_x,d_y\le u.
\end{equation}

According to \cite{weisstein2004square}, we can calculate the project distance between Alice and Willie in the following way. Since $d_{aw}^{\prime}=\sqrt{(u_{ax}-u_{wx})^2+(u_{ay}-u_{wy})^2}$, we consider the Polar Coordinate and let $|u_{ax}-u_{wx}|=d \cos{\theta}$ and $|u_{ay}-u_{wy}|=d \sin{\theta}$, where $0\le d\le u\sqrt{2}$ and $1\le\theta\le\frac{\pi}{2}$. According to \eqref{eq:trian}, the joint PDF can be recast to

\begin{equation}
    \label{eq:distance1}
    f(d,\theta)=\frac{4d(u-d \cos{\theta})(u-d \sin{\theta})}{u^4}.
\end{equation}

When $0\le d\le u$, the marginal PDF of \eqref{eq:distance1} as a function of $d$ can be written as
\begin{equation}
    \begin{aligned}
        f_D(d)&=\frac{4r}{u^4}\int_0^{\frac{\pi}{2}}(u-d \cos{\theta})(u-d \sin{\theta})\mathrm{d}\theta\\
        &=\frac{2\pi d}{u^2}-\frac{8d^2}{u^3}+\frac{2d^3}{u^4}.
    \end{aligned}
\end{equation}

Similarly, when $u<d\le u\sqrt{2}$, according to the limitations $d \cos{\theta}\le u$ and $d \sin{\theta}\le u$, we have $\theta\in[\arccos{\frac{u}{d}}, \arcsin{\frac{u}{d}}]$. The marginal PDF of \eqref{eq:distance1} as a function of $d$ can be written as

\begin{equation}
    f_D(r)=\frac{2d}{u^2}(\pi-\frac{4d}{u}+\frac{5d^2}{u^2}-4-\frac{4\sqrt{d^2-u^2}}{u}).
\end{equation}

Therefore, the PDF of the projected distance $d_{aw}^{\prime}$ can be expressed as

\begin{equation}
\label{eq:distance}
 f_D(d)=
~\begin{cases} 
\frac{2d}{u^2}(\pi -\frac{4d}{u}+\frac{d^2}{u^2}), & 0\le d\le u, \\ 
\frac{2d}{u^2}(\frac{4}{u}\sqrt{d^2-u^2}-(\frac{d^2}{u^2}+2-\pi)\\-4\arctan \frac{\sqrt{d^2-u^2}}{u}), & u<d\le u\sqrt{2}.
\end{cases}   
\end{equation}

Then the cumulative distribution function (CDF) of $ f_D(d)$ can be expressed as
\begin{equation}
\label{eq:CDF}
    P_{CDF}(d)=
    ~\begin{cases} 
\frac{d^4}{2u^4}-\frac{8d^3}{3u^3}+\frac{\pi d^2}{u^2}, & 0\le d\le u, \\ 
\frac{1}{3}-\frac{d^4}{2u^4}-\frac{4d^2\arctan \frac{\sqrt{d^2-u^2}}{u}}{u^2}\\+\frac{4}{3}(\frac{2d^2}{u^2}+1)\frac{\sqrt{d^2-u^2}}{u}+\frac{(\pi-2)d^2}{u^2}, & u<d\le u\sqrt{2}.
\end{cases}  
\end{equation}

Furthermore, we find the expectation of $ f_D(d)$  as
\begin{equation}
\label{eq:Expect}
   \expec_{f_D}(d)=
    ~\begin{cases} 
\frac{2d^5}{5u^4}-\frac{2d^4}{u^3}+\frac{2\pi d^3}{3u^2}, & 0\le d\le u, \\ 
\frac{2u}{15}-\frac{2d^5}{5u^4}-\frac{8d^3\arctan \frac{\sqrt{d^2-u^2}}{u}}{3u^2}\\+(\frac{2d^3}{u^2}+\frac{1}{3}d)\frac{\sqrt{d^2-u^2}}{u}\\+\frac{u}{3}\ln{(\frac{d}{u}+\frac{\sqrt{d^2-u^2}}{u})}+\frac{2(\pi-2)d^3}{3u^2}, & u<d\le u\sqrt{2}.
\end{cases}  
\end{equation}

Finally, the distance between Alice and Willie $d_{aw}$ can be expressed as 
\begin{equation}
\label{eq:gougu}
  d_{aw}=\sqrt{d^{\prime 2}_{aw}+H_{w}^{2}}.
\end{equation}
Where $H_w$ is the height of the UAV Willie.

\subsection{Catching Probability and Optimal Detection Window Size}
We assume that Alice needs to transmit the message with the length of $M$ bits per Herz. The unit transmission rate $R_{ab}$ can be calculated from \eqref{Eq:rate} $R_{ab}$. Then, the Monte Carlo simulation calculates the average transmission rate $\overline{R}_{ab}$. As we mentioned in the system model, to focus on the problem, we do not consider the corresponding channel codes for the bit rate error and the modulation techniques to simplify the system model. We assume that Alice transmits one bit in each symbol. Thus, we find Alice's transmission time $t_{tr}$ is
\begin{equation}
    t_{tr}=\frac{M}{\overline{R}_{ab}}.
\end{equation}

Willie can catch Alice only when he detects the transmission and locates Alice by chasing the signal within time $t_{tr}$. Here, we assume that Willie is not always detecting the transmission. Instead, he will wait for the transmission time of $l_s$ symbols after the first detection and collect the following $L$ symbols to detect. Therefore, during transmission, Willie can detect the transmission multiple times. Meanwhile, when Willie is far from Alice, even though he detects the transmission, he may not be able to find Alice during the transmission. Thus, we consider using the effective detection frequency $s$ to measure the number of detections Willie can make to find Alice.

The probability of catching is determined by the projected distance $d_{aw}^{\prime}$, the projected radius $r_a^{\prime}$, and $r_w^{\prime}$.
When $d_{aw}^{\prime}\le r_a^{\prime}$, Alice can observe Willie. Thus, she will not transmit the message to the LEO satellite Bob, and the catching probability is $P_{ca}=0$. 

When $r_a^{\prime}\le d_{aw}^{\prime}\le r_w^{\prime}$, Willie can find Alice immediately when he detects the transmission without being observed by Alice, and the maximum effective detection frequency $s_m$ is given by 
\begin{equation}
\label{eq:maxs}
    s_m=\lceil\frac{M-L}{L+l_s}\rceil.
\end{equation}
Where $\lceil z\rceil$ is the ceil of $z$. 
Then, the catching probability under the condition that the effective detection frequency is $s_m$ is
\begin{equation}
\label{eq:catch1}
    P_{ca}(L,d_{aw}|s=s_m)=1-(P_{MD}(L,d_{aw}))^{s_m}.
\end{equation}
Where $P_{MD}(L,d_{aw})$ is the miss-detection probability in \eqref{eq:MD} as a function of $L$ and $d_{aw}$. 

When $d_{aw}^{\prime}>r_w^{\prime}$, if $\frac{M-L}{L+l_s}$ is not an integer, the catching probability in \eqref{eq:catch1} holds for the following range.
\begin{equation}
\begin{aligned}
     &s_m-1\le\frac{M-L-\frac{(d^{\prime}_{aw}-r_{w}^{\prime})\overline{R}_{ab}}{v_w}}{L+l_s}\\
     \Leftrightarrow & d^{\prime}_{aw}\le\frac{v_w(M-L-(L+l_s)(s_m-1))}{\overline{R}_{ab}}+r_{w}^{\prime}.
\end{aligned}
\end{equation}

Therefore, when $r_a^{\prime}\le d^{\prime}_{aw}\le \frac{v_w(M-(L+l_s)s_m)}{R_{ab}}+r_{w}^{\prime}$, the conditional catching probability can be calculated from  \eqref{eq:catch1}.

When the effective detection frequency is $s=i$ ( where $1\le i< s_m$), we have

\begin{equation}
\label{eq:range_s}
\begin{aligned}
     &i-1\le\frac{M-L-\frac{(d^{\prime}_{aw}-r_{w}^{\prime})\overline{R}_{ab}}{v_w}}{L+l_s}<i\\
     \Leftrightarrow & d^{\prime}_{aw}>\frac{v_w(M-L-(L+l_s)i)}{\overline{R}_{ab}}+r_{w}^{\prime}  ~~\text{and}~~\\
     &d^{\prime}_{aw}\le\frac{v_w(M-L-(L+l_s)(i-1)}{\overline{R}_{ab}}+r_{w}^{\prime}.
\end{aligned}
\end{equation}
Then the catching probability under the condition that the effective detection frequency is $s=i$ is
\begin{equation}
\label{eq:catch2}
    P_{ca}(L,d_{aw}|s=i)=1-(P_{MD}(L,d_{aw}))^{i}.
\end{equation}

Since the distance $d^{\prime}_{aw}$ is a random variable with the probability density function $f_D(d)$ in \eqref{eq:distance}, the probability of the corresponding distance range can be calculated from \eqref{eq:CDF}.
\begin{equation}
\label{eq:PD}
    P_{dis}(s=i)=\int_{d_1}^{d_2}f_D(d)\mathrm{d}d=P_{CDF}(d_2)-P_{CDF}(d_1).
\end{equation}
Here $[d_1,d_2]$ is the range of $d^{\prime}_{aw}$ previous introduced.

To find the expression of the catching probability $P_{ca}(L)$ as a function of $L$, we find the approximation results for \eqref{eq:catch1} and \eqref{eq:catch2}. According to \eqref{eq:Expect}, the expectation of $d^{\prime}_{aw}$ in range $[d_1,d_2]$ can be calculated by
\begin{equation}
\label{eq:expectdis}
    \overline{d}^{\prime}_{aw}=\expec_{f_D}(d_2)-\expec_{f_D}(d_1)+d_1.
\end{equation}

From \eqref{eq:gougu}, the average distance between Alice and Willie is $\overline{d}_{aw}=\sqrt{\overline{d}^{\prime 2}_{aw}+H_w^2}$. Therefore, for $1\le i\le s_m$, the approximated catching probability under the condition $s=i$ is
\begin{equation}
\label{eq:Approx}
    P_{ca}(L, \overline{d}_{aw}|s=i)=1-(P_{MD}(L, \overline{d}_{aw}))^{i}.
\end{equation}

Finally, according the \eqref{eq:PD} and \eqref{eq:Approx}, the catching probability can be calculated by
\begin{equation}
\label{eq:catch}
    P_{ca}(L)=\sum_{i=1}^{s_m}P_{ca}(L, \overline{d}_{aw}|s=i)P_{dis}(s=i).
\end{equation}

According to the above formulas, we propose Algorithm~\ref{Alg:1} to solve the optimization problem $\max_{L} P_{ca}(L)$. 

\begin{algorithm}
\caption{Optimal Detection Window Size}
\label{Alg:1}
\begin{algorithmic}[1]
\Require Channel gains $G_a$, $G_w$, $\eta$, square area size $u$, speed $v_w$, message length $M$, detection interval $l_s$, noise power $\sigma_w^2$, transmission power $P_a$, projected radius $r_a^{\prime}$, $r_w^{\prime}$, average transmission rate $\overline{R}_{ab}$
\Ensure Optimal detection window size $L^*$
\For{$j \gets1  ~\text{to}~ M$}
\State $\Gamma_w^{(j)} \gets \delta=1-\frac{\gamma(j,\frac{j\Gamma_w}{\sigma_w^2})}{\Gamma(j)}$ 
\State $s_m^{(j)} \gets \lceil\frac{M-j}{j+l_s}\rceil$
\For{$i \gets 1  ~\text{to}~ s_m^{(j)}$}
\If{$i==s_m^{(j)}$}
\State $d_1\gets r_a$ 
\State $d_2 \gets \frac{v_w(M-j-(j+l_s)(s_m^{(j)}-1))}{\overline{R}_{ab}}+r_{w}^{\prime}$
\ElsIf{$i<s_m^{(j)}$}
\State $d_1\gets \frac{v_w(M-j-(j+l_s)i)}{\overline{R}_{ab}}+r_{w}^{\prime}$
\State $d_2 \gets \frac{v_w(M-j-(j+l_s)(i-1))}{\overline{R}_{ab}}+r_{w}^{\prime}$
\EndIf
\State $P_{dis}(s=i)=P_{CDF}(d_2)-P_{CDF}(d_1)$
\State $\overline{d}^{\prime}_{aw}=\expec_{f_D}(d_2)-\expec_{f_D}(d_1)+d_1$
\State $\overline{g}_{aw}\gets\frac{\sqrt{\eta G_a G_w}}{(\overline{d^{\prime}}_{aw}^{2}+H_{w}^{2})^{\alpha_N/2}}$
\State $\overline{P}_{MD}\gets \frac{\gamma(j,\frac{j\Gamma_w^{(j)}}{ \overline{g}_{aw}^2P_a+\sigma_w^2})}{\Gamma(j)}$
\State $P_{ca}(j|s=i)\gets 1-(\overline{P}_{MD})^{i}$
\EndFor
\State $P_{ca}(j)\gets \sum_{i=1}^{s_m^{(j)}}P_{ca}(j|s=i)P_{dis}(s=i)$
\EndFor
\State $L^*=\arg\max\limits_{ j\in [1,M]} P_{ca}(j)$
\end{algorithmic}
\end{algorithm}

\section{Message Splitting at Alice}
\label{sec:alice}
In this section, we mainly derive the expression of the overall catching probability when applying the message splitting at Alice. Then, we analyze how the overall catching and catching probabilities change with system parameters such as Alice's transmission power, message length, and square area size. Finally, we propose two cases to study our system model further.

\subsection{Overall Catching Probability}
From the \eqref{eq:maxs}, we know that the effective detection frequency increases with the message length $M$, which leads to an increase in catching probability for each chunk according to \eqref{eq:catch2} and \eqref{eq:catch}. 
Therefore, Alice may consider splitting the message with length $M$ into $n$ smaller chunks with length $M/n$. Thus, Willie's catching probability for each chunk will decrease accordingly. However, Alice needs to transmit the whole message with $n$ transmission slots. In other words, Willie will have a $n-1$ higher chance of detecting the transmission of the message. Whether the message splitting method will decrease the overall catching probability for a message needs further study. Here, we will derive the overall catching probability expression similar to the previous section's derivations.

The maximum effective detection frequency is $s_m=\lceil\frac{M/n-L}{L+l_s}\rceil$. 
When $d_{aw}\le\frac{v_w(M/n-L-(L+l_s)(s_m-1))}{\overline{R}_{ab}}+r_{w}$, according to \eqref{eq:catch1} and \eqref{eq:expectdis}, the approximated catching probability for a chunk under the condition that the effective detection frequency $s=s_m$ can be e expressed as
\begin{equation}
\label{eq:catchA1}
    P_{ca}(n,\overline{d}_{aw}|s=s_m)=1-(P_{MD}(\overline{d}_{aw})^{s_m}.
\end{equation}
Notice that the detection window size $L$ is a given parameter and $P_{MD}(\overline{d}_{aw})$ is a function of $\overline{d}_{aw}$.

When the effective detection frequency is $s=i$ ( where $1\le i< s_m$), according to \eqref{eq:range_s}, the projected distance $d_{aw}^{\prime}$ satisfies  $d_{aw}^{\prime}>\frac{v_w(M/n-L-(L+l_s)i)}{\overline{R}_{ab}}+r_{w}^{\prime}$ and $d_{aw}^{\prime}\le\frac{v_w(M-L-(L+l_s)(i-1))}{\overline{R}_{ab}}+r_{w}^{\prime}$.
Then the approximated catching probability for a chunk under the condition that the effective detection frequency is $s=i$ is given by
\begin{equation}
\label{eq:catchA2}
    P_{ca}(n|s=i)=1-(P_{MD}(\overline{d}_{aw}))^{i}.
\end{equation}

According to \eqref{eq:PD}, the catching probability for a chunk can be expressed as
\begin{equation}
\label{eq:catchA}
    P_{ca}(n)=\sum_{i=1}^{s_m}P_{ca}(n,\overline{d}_{aw}|s=i)P_{dis}(s=i).
\end{equation}

Then, if Willie does not catch Alice, the probability will be $P_{nc}(n)=1-P_{ac}(n)$. In a new transmission slot, Alice will observe Willie and stop transmitting if he is located within Alice's $r_a$ radius. Thus, the transmission is postponed to the next transmission slot; that is, Alice needs to spend one more transmission slot. This scenario will happen with the probability $P_{as}$, which is given by

\begin{equation}
    P_{as}=\int_{0}^{r_a}f_D(d)\mathrm{d}d=P_{CDF}(r_a).
\end{equation}

Alice must spend at least $n$ times to complete transmitting the whole message. If Alice spends $n$ times, the probability of not being caught by Willie is $(P_{nc}(n)-P_{as})^n$. Then, if Alice spends $n+1$ times, the probability is $\binom{1}{n}(P_{nc}(n)-P_{as})^nP_{as}$. Similarly, if Alice spends $n+2$ times, the probability is $\binom{2}{n+1}(P_{nc}(n)-P_{as})^nP_{as}^2$. Continuously, if Alice spends $n+i$ times, the probability is $\binom{i}{n+i-1}(P_{nc}(n)-P_{as})^nP_{as}^i$. Therefore, the sum of all the probabilities is 
\begin{equation}
\label{eq:mid}
\begin{aligned}
    P &= \sum_{i=0}^{\infty}\binom{i}{n+i-1}(P_{nc}(n)-P_{as})^nP_{as}^i\\
    &=(P_{nc}(n)-P_{as})^n\sum_{i=0}^{\infty}\binom{i}{n+i-1}P_{as}^i.
    \end{aligned}
\end{equation}

According to the negative binomial theorem, the probability in \eqref{eq:mid} can be recast to
\begin{equation}
    P= (\frac{1-P_{ca}(n)-P_{as}}{1-P_{as}})^n.
\end{equation}

Therefore, the overall catching probability for a message can be expressed as
\begin{equation}
\label{eq:ov}
    \begin{aligned}
     P_{ov}(n)=1-(\frac{1-P_{ca}(n)-P_{as}}{1-P_{as}})^n.
     \end{aligned}
\end{equation}

According to the above formulas, we propose Algorithm~\ref{Alg:2} to solve the optimization problem $\min_{n} P_{ov}(n)$.

\begin{algorithm}
\caption{Optimal Splitting-Chunks Quantity}
\label{Alg:2}
\begin{algorithmic}[1]
\Require Channel gains $G_a$, $G_w$, $\eta$, square area size $u$, speed $v_w$, message length $M$, detection interval $l_s$, noise power $\sigma_w^2$, transmission power $P_a$, projected radius $r_a^{\prime}$, $r_w^{\prime}$, average transmission rate $\overline{R}_{ab}$, detection window size $L$
\Ensure Optimal splitting-chunks quantity $n^*$
\State $\Gamma_w \gets \delta=1-\frac{\gamma(L,\frac{L\Gamma_w}{\sigma_w^2})}{\Gamma(L)}$ 
\State $P_{as}\gets P_{CDF}(r_a)$
\For{$j \gets1  ~\text{to}~ N$}
\State $s_m^{(j)} \gets \lceil\frac{M/j-L}{L+l_s}\rceil$
\For{$i \gets 1  ~\text{to}~ s_m^{(j)}$}
\If{$i==s_m^{(j)}$}
\State $d_1\gets r_a$ 
\State $d_2 \gets \frac{v_w(M/j-L-(L+l_s)(s_m^{(j)}-1))}{\overline{R}_{ab}}+r_{w}^{\prime}$
\ElsIf{$i<s_m^{(j)}$}
\State $d_1\gets \frac{v_w(M/j-L-(L+l_s)i)}{\overline{R}_{ab}}+r_{w}^{\prime}$
\State $d_2 \gets \frac{v_w(M/j-L-(L+l_s)(i-1))}{\overline{R}_{ab}}+r_{w}^{\prime}$
\EndIf
\State $P_{dis}(s=i)=P_{CDF}(d_2)-P_{CDF}(d_1)$
\State $\overline{d}^{\prime}_{aw}=\expec_{f_D}(d_2)-\expec_{f_D}(d_1)+d_1$
\State $\overline{g}_{aw}\gets\frac{\sqrt{\eta G_a G_w}}{(\overline{d^{\prime}}_{aw}^{2}+H_{w}^{2})^{\alpha_N/2}}$
\State $\overline{P}_{MD}\gets \frac{\gamma(L,\frac{L\Gamma_w}{ \overline{g}_{aw}^2P_a+\sigma_w^2})}{\Gamma(L)}$
\State $P_{ca}(j|s=i)\gets 1-(\overline{P}_{MD})^{i}$
\EndFor
\State $P_{nc}(j)\gets 1- \sum_{i=1}^{s_m^{(j)}}P_{ac}(j|s=i)P_{dis}(s=i)$
\State $P_{ov}(j)\gets 1-(\frac{P_{nc}-P_{as}}{1-P_{as}})^j$
\EndFor
\State $n^*=\arg\min\limits_{ j\in [1,N]} P_{ov}(j)$
\end{algorithmic}
\end{algorithm}

Although Alice splits the message into smaller sizes, the split message length is generally much larger than the detection window size, $M/n> L$. This is because the splitting-chunks quantity $n$ are always constrained by $N$. Two reasons determine the constraint. First, when Alice decides to transmit $n$ chunks separately for each message, she needs to spend $n$ transmission slots. The time $t_c$ spent in each transmission slot is usually much larger than the transmission time. Therefore, the time cost may not be acceptable with a large $n$. 
Second, Alice's transmission system may not be stable with a large $n$. For example, assuming Alice will receive the message follows the Poisson process with rate $\lambda$. We can approximate the system as an M/G/1 queue. Since $t_c$ is constant, the execution time can be approximated to $nt_c$. Therefore, the system is stable only when $n\le \lambda t_c$.

\subsection{System Parameters Analysis}
\label{subsec:parameter}

According to the system model, many system parameters can affect the probability of catching and the overall probability of catching. As mentioned, we mainly focus on the detection window size $L$ and the splitting-chunks quantity $n$. Other parameters are also worth studying. In this paper, we will select the other three parameters to analyze: Alice's transmission power $P_a$, the message length $M$, and the square area size $u$.
\subsubsection{Alice's transmission power $P_a$}
From \eqref{Eq:rate}, we know that the transmission rate increases with $P_a$, which leads to Willie having less time to locate Alice during the transmission. Meanwhile, from \eqref{eq:MD}, the probability of missed detection decreases with increasing $P_a$. Thus, it is unclear how $P_a$ affects catching and its overall probability.

\subsubsection{Message length $M$}
From \eqref{eq:range_s} and \eqref{eq:PD}, we know that the effective detection frequency increases with $M$, which leads to Willie having more chances to catch Alice. From \eqref{eq:catchA1}, the overall catching probability also increases with $M$.  

\subsubsection{Square area size $u$}
From \eqref{eq:gain_aw}, we know that the miss detection probability increases with the distance between Alice and Willie.
The average distance will also increase with the increase of $u$, according to \eqref{eq:PD}. Thus, the catching and the overall catching probabilities decrease with $u$.

\subsection{Cases Study}
\label{subsec:case}
Previously, we have proposed Algorithm~\ref{Alg:2} to find the optimal splitting-chunks quantity $n$. However, the algorithm only considers evaluating the overall catching probability with one message. However, Alice may need to transmit multiple messages in a period. In practice, Alice and Willie cannot work for an infinite time due to the energy constraint; then, we assume they will continuously work for $\beta$ transmission slots in the following two cases. We also assume the messages that arrive at Alice follow the Poisson process. Here, we consider using the number of covert messages transmitted during the $\beta$ transmission slots as the new performance metric. We define the covert transmitted messages as Alice finishes transmitting the message to Bob without being caught by Willie.

\subsubsection{Case 1} In this case, we consider that Alice can transmit the message covertly only if Willie never catches her. Once Willie catches Alice, she loses the chance to transmit covertly in the rest of the transmission slots. Thus, Alice prioritizes reducing the overall catching probability in determining the optimal $n$, and the constraint $N$ will be ignored.

\subsubsection{Case 2} In this case, we consider that only the current transmission is detected when Willie catches Alice. Alice can still transmit the following message covertly. Thus, the number of messages transmitted within $\beta$ time intervals is also essential to performance. Alice may prefer a smaller $n$ to maximize the number of covered messages transmitted.

\section{Simulation Results and Discussion}
\label{sec:simulation}

\begin{table}[hbt]
\begin{center}
\begin{threeparttable}
    \caption{Simulation Parameters }
    \label{Table:sim}
				 \begin{tabular}{c|c}
    \toprule
    {\bf Parameters} & {\bf Values}\\ \hline  &\\[-2.5mm] 
          Alice's transmit power & $P_a=0 $ dB \\
          &\\[-3mm] 
       \hline 
       &\\[-2.5mm] 
         Antenna gain & $G_a = 0$ dB, $G_b=18$ dB, $G_w=-40$ dB \\
         &\\[-2.5mm] 
          \hline
           Channels' coefficients & $K_0=5$,$\eta=-10$ dB\\ \hline &\\[-2.5mm]
       Message length & $M=600$ bits\\ \hline &\\[-2.5mm]      
      Noise power & $\sigma^2_w=-90$ dBm, $\sigma^2_b=-110$ dBm\\ \hline &\\[-2.5mm] 
      False alarm constraint & $\delta=0.05$\\ \hline &\\[-2.5mm]
      Detection interval & $l_s=20$ symbols\\ \hline &\\[-2.5mm] 
      Square area size & $u=1000$ m\\ \hline &\\[-2.5mm] 
      LEO satellite distance & $d_{ab}=500$ km\\ \hline &\\[-2.5mm] 
       UAV flying height & $H_w=150$ m\\ \hline
         &\\[-2.5mm]  
      UAV moving speed & $v_{w}=15$ m/s\\ \hline &\\[-2.5mm] 
      Alice's observation radius & $r_a^{\prime}=100$ m\\ \hline &\\[-2.5mm] 
      Willie's catching radius & $r_w^{\prime}=150$ m\\
       \bottomrule
    \end{tabular}
    \end{threeparttable}
	\end{center}
\end{table}
This section presents simulation results to evaluate how the catching probability changes with different system parameters and verify the effectiveness of the proposed algorithms.
Unless otherwise stated, the simulation parameters can be found in Table~\ref{Table:sim}. 
Notice that the curves in the following simulation results are not smooth enough because the related parameters, such as detection window size, splitting-chunks quantity, and message length, are integers. 

\subsection{Detection Window Size Analysis for Willie}
First, we consider Willie's perspective. We will analyze how to select the detection window size to increase the catching probability. Meanwhile, we also analyze three other vital system parameters: Alice's transmission power, the message length, and the square area size.  

\begin{figure}[htb]
    \centering
    \includegraphics[width=0.5\textwidth]{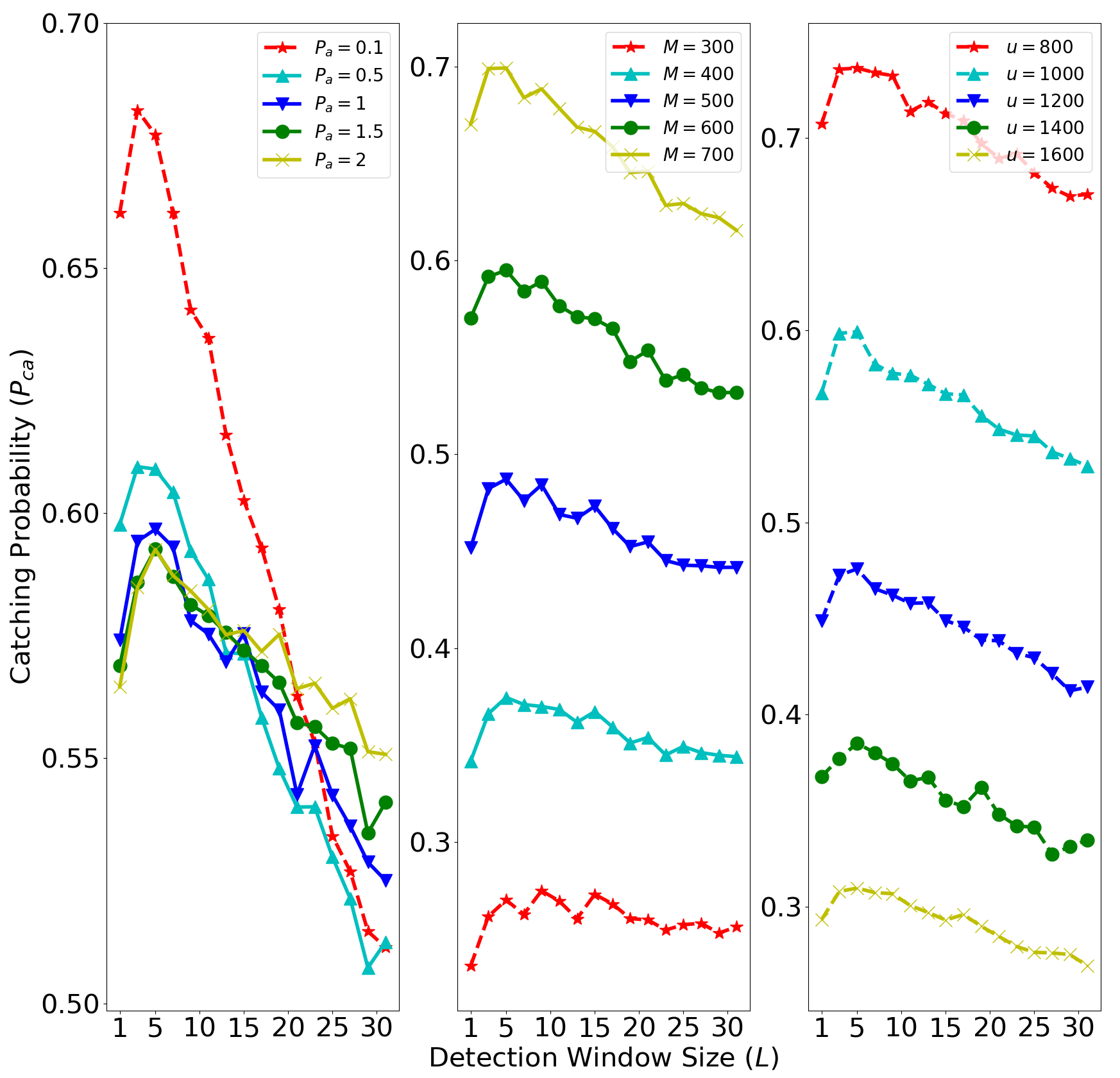}
    \caption{The catching probability $P_{ca}$ versus the detection window size $L$ for Alice's transmission power $P_{a}$ (left), the message length $M$ (middle), and the square area size $u$ (right), respectively.   }
    \label{fig:change_L}
\end{figure}

In Figure~\ref{fig:change_L}, we evaluate the catching probability $P_{ca}$ versus the detection window size $L$ for different system parameters. In the left graph, we consider Alice's transmission power $P_{a}$ for five different values in $\{0.1, 0.5, 1, 1.5, 2\}$. In the middle graph, we consider the message length $M$ for five different values in $\{300, 400, 500, 600, 700\}$. In the right graph, we consider the square area size $u$ for five different values in $\{800, 1000, 1200, 1400, 1600\}$. 

All three graphs show that the detection window size can affect the catching probability by $5\%$ to $15\%$. We can observe that when the detection window size is smaller, the catching probability increases with $L$; when the detection window size is large, the catching probability decreases with increasing $L$. Therefore, there is an optimal $L$ that maximizes the catching probability. The figure also shows that the optimal $L$ is generally small, such as $L^*\le 10$. The observation is consistent with the results shown in \eqref{eq:MD} and \eqref{eq:Approx}. Although increasing $L$ can always reduce the miss detection probability $P_{MD}$, the improvement is trivial when $P_{MD}$ reaches the upper limit. Then, continuously increasing $L$ will reduce the effective detection infrequency, which leads to a decrease in the catching probability.

From Figure~\ref{fig:change_L}, we can also observe how the other three system parameters affect the catching probability. In the left graph, $P_{ca}$ decreases with the increasing $P_a$ when $L$ is small, and  $P_{ca}$ increases with $P_a$ when $L$ is large. The observations are consistent with our analysis in Section~\ref{subsec:parameter}. Although increasing the transmission power can reduce Willie's miss detection probability, Alice can transmit the message faster, so there is less time left for Willie to locate Alice.
The middle graph shows that $P_{ca}$ increases with $M$. When the message length is large, Alice needs to transmit more bits over a longer time. Then Willie has more chances to detect the transmission and locate Alice. In the right graph, $P_{ca}$ decreases with the increasing $u$. As the increase of $u$, Willie is responsible for a larger area, but his detection ability does not improve. 

\begin{figure}[htb]
    \centering
    \includegraphics[width=0.5\textwidth]{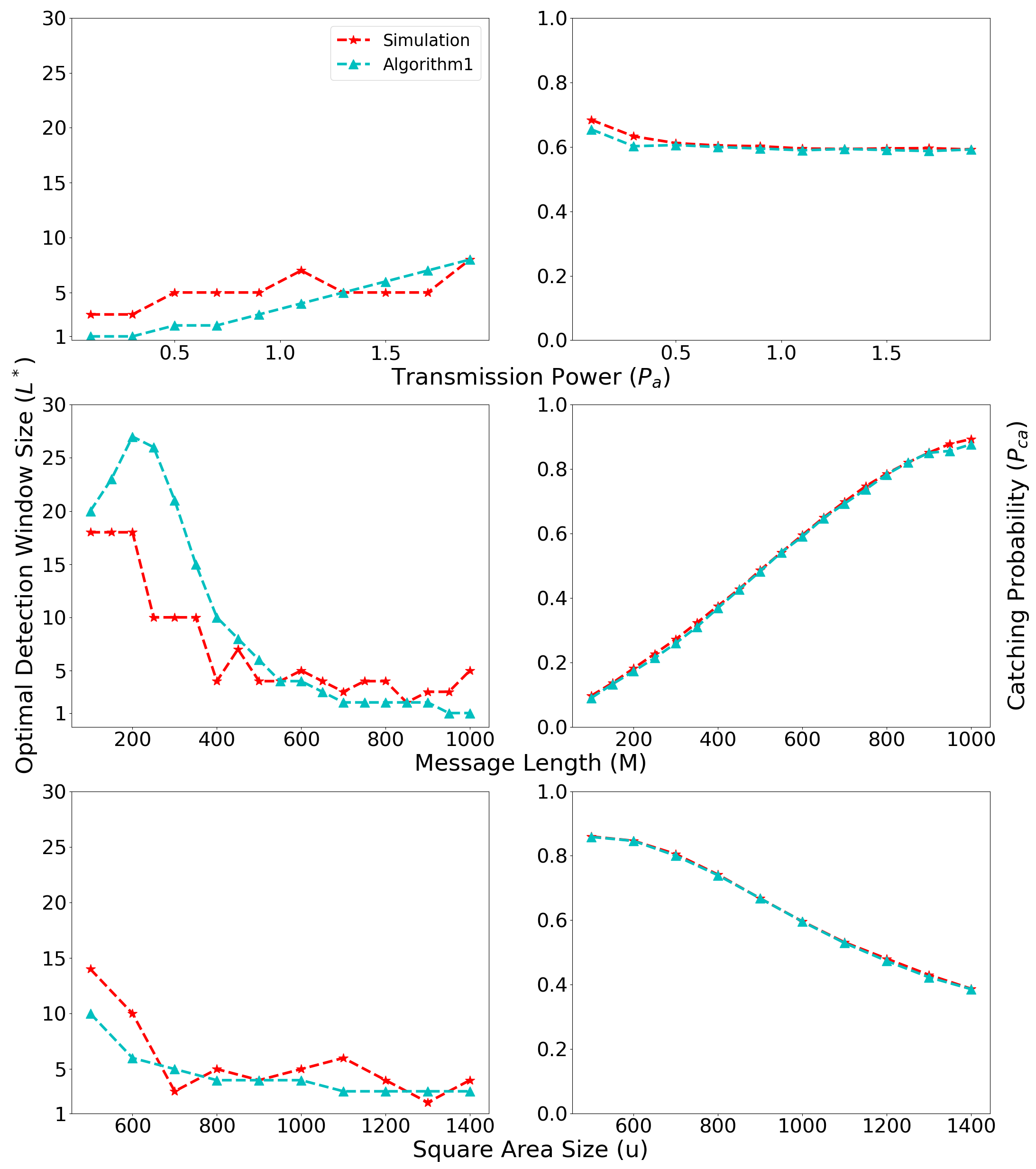}
    \caption{The optimal detection window size $L^*$ (left) and its corresponding catching probability $P_{ca}$ (right) versus the Alice's transmission power $P_{a}$ (upper), the message length $M$ (middle), and the square area size $u$ (lower), respectively. }
    \label{fig:opt_L}
\end{figure}

In Figure~\ref{fig:opt_L}, we separately evaluate how the optimal detection window size $L^*$ changes with Alice's transmission power $P_{a}$, the message length $M$, and the square area size $u$ in the three graphs in the left. Meanwhile, we provide the corresponding catching probability $P_{ca}$ for each value of $L^*$. In each graph, we also evaluate the performance of Algorithm~\ref{Alg:1} by comparing it with the simulation.
The upper two graphs show that the detection window size generally increases slowly with $P_{a}$, and the corresponding catching probability also slowly decreases with the increasing $P_{a}$. The phenomenon is similar to the observation in Figure~\ref{fig:change_L}. The middle two graphs show that $L^*$ generally decreases with the increasing $M$ and $P_{ca}$ increases with $M$. The observation indicates that when Willie has more chances to detect the transmission as the message length increases, it is better to sacrifice the miss detection probability and increase the effective detection frequency. The lower graph shows that $L^*$ and $P_{ca}$ generally decrease with the increasing $u$. The reason is that although Willie detects the transmission with higher probability by increasing $L$ as the increase of $u$, he may not have enough time to chase Alice during the transmission. Figure~\ref{fig:opt_L} also shows that although Algorithm~\ref{Alg:1} does not fit the simulation perfectly, the corresponding catching probability almost coincides with the simulation results because the detection window size is not very sensitive to the catching probability, which we can observe from Figure~\ref{fig:change_L}.

\subsection{Splitting-Chunks Quantity Analysis for Alice}
Second, we consider Alice's perspective. We will analyze how to select the quantity of splitting chunks to decrease the catching probability. Meanwhile, we also analyze three other vital system parameters: Alice's transmission power, the message length, and the square area size.  

\begin{figure}[htb]
    \centering
    \includegraphics[width=0.5\textwidth]{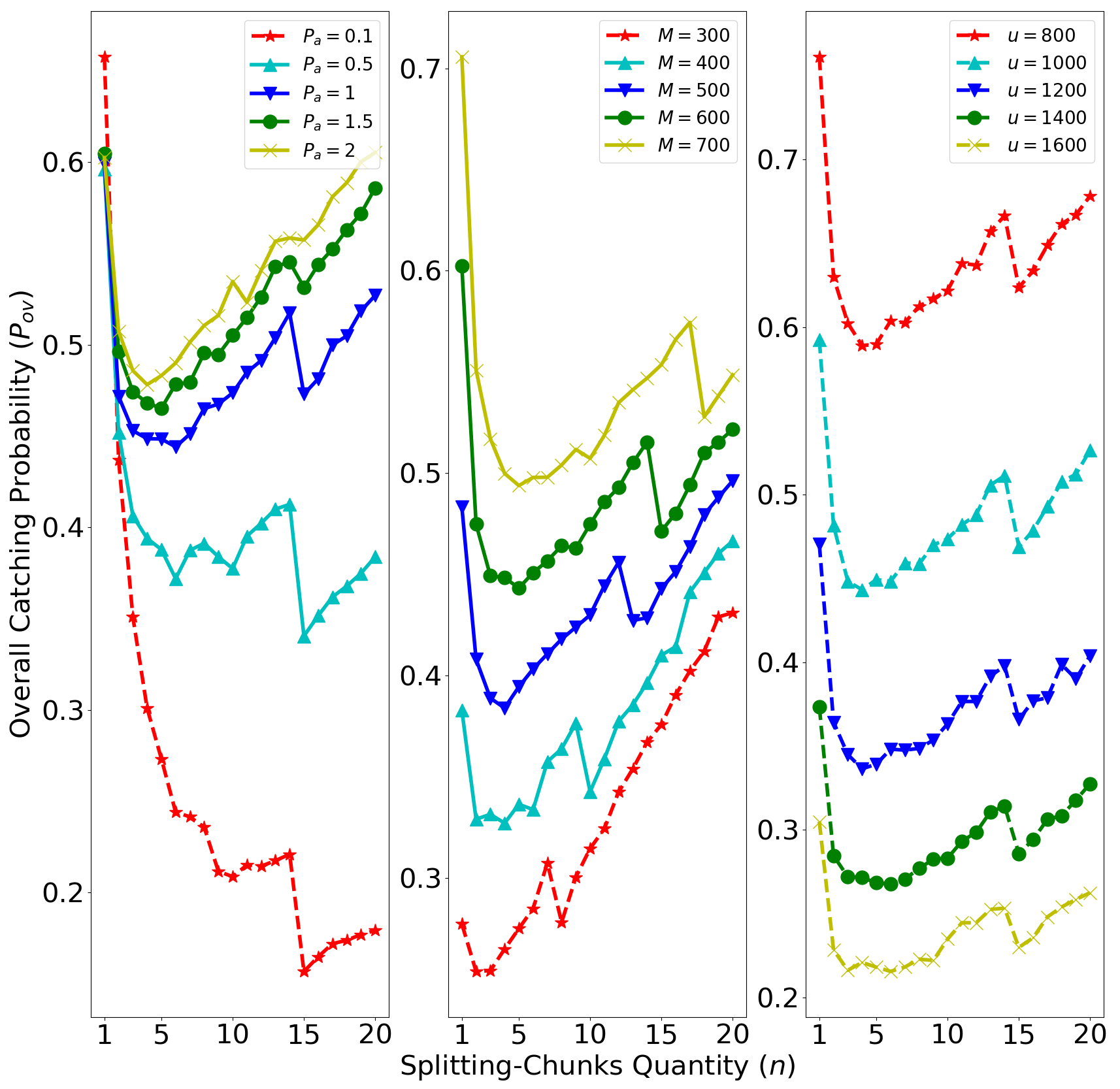}
    \caption{The overall catching probability $P_{ov}$ versus the splitting-chunks quantity $n$ for Alice's transmission power $P_{a}$ (left), the message length $M$ (middle), and the square area size $u$ (right), respectively. }
    \label{fig:change_n}
\end{figure}

In Figure~\ref{fig:change_n}, we evaluate the overall catching probability $P_{ov}$ versus the splitting-chunks quantity $n$ for different system parameters. We consider the detection window size $L=10$. In the left graph, we consider Alice's transmission power $P_{a}$ for five different values in $\{0.1, 0.5, 1, 1.5, 2\}$. The middle graph considers the message length $M$ for five different values in $\{300, 400, 500, 600, 700\}$. The right graph considers the square area size $u$ for five different values in $\{800, 1000, 1200, 1400, 1600\}$. 

All three graphs show that the quantity of splitting-chunks can significantly affect the overall catching probability by about $20\%$. 
We can observe that the overall catching probability decreases rapidly with the increasing $n$ when $n$ is small, and increases slowly with $n$ when $n$ is large. Splitting the message aims to reduce the catching probability for each chunk by providing more transmissions for detection. The method performs well when the reduction is significant. According to the middle graph of Figure~\ref{fig:change_L}, $P_{ca}$ almost decreases linearly with the message length. Therefore, $P_{ov}$ can be reduced significantly. However, when $n$ is large, the catching probability for signal detection will be reduced to a lower limit. Then, increasing $n$ will lose its benefits.
We can observe how the other three system parameters affect the overall catching probability. The observations are similar to the results in Figure~\ref{fig:change_L} because the overall catching probability $P_{ov}$ has similar properties to $P_{ca}$ when $n$ is given according to \eqref{eq:ov}. Besides, we also observe that the optimal $n$ changes significantly with $P_{a}$ and $M$; however, it almost remains the same considering $u$.

\begin{figure}[htb]
    \centering
    \includegraphics[width=0.5\textwidth]{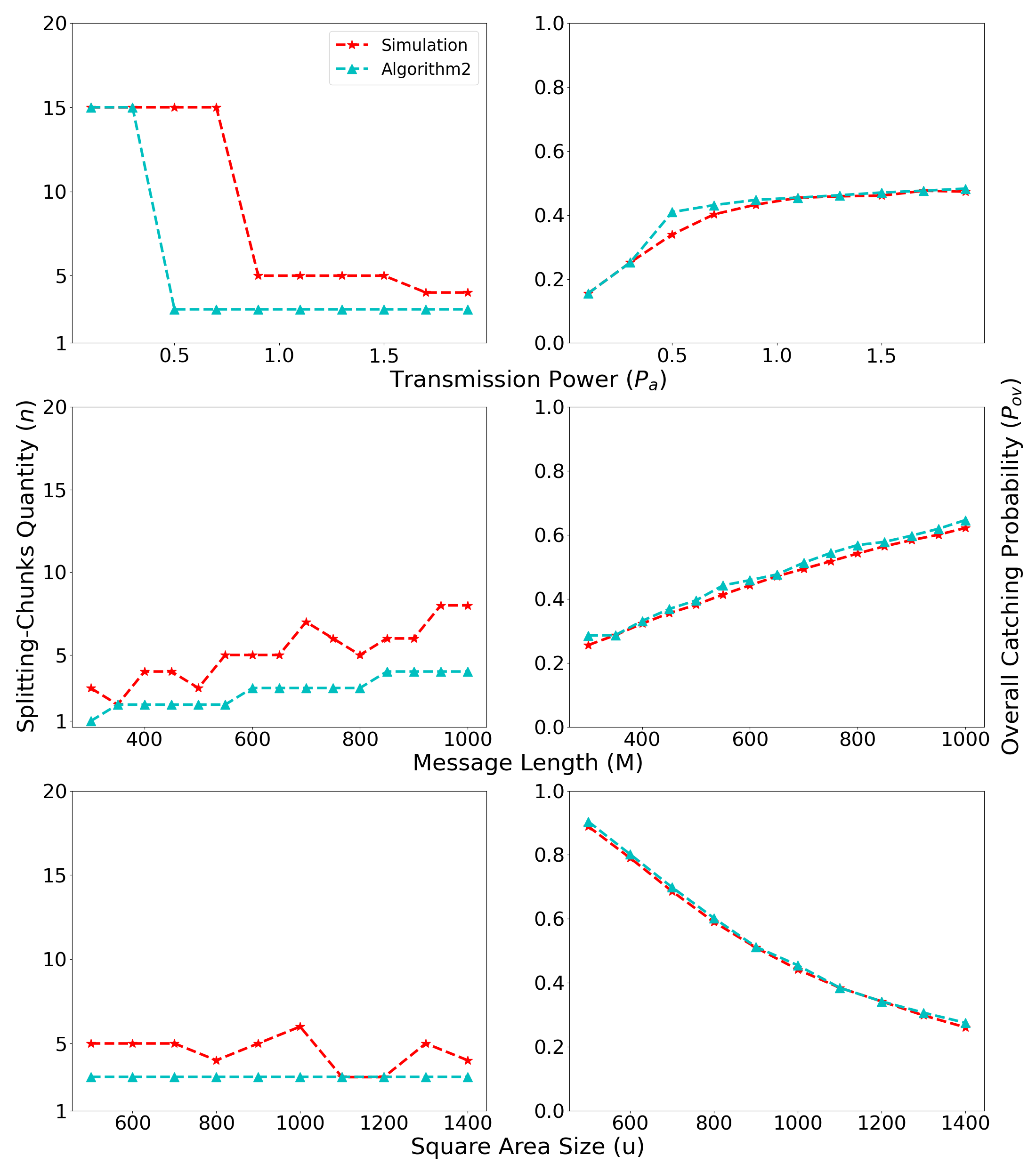}
    \caption{The optimal splitting-chunks quantity $n^*$ (left) and its corresponding overall catching probability $P_{ov}$ (right) versus the Alice's transmission power $P_{a}$ (upper), the message length $M$ (middle), and the square area size $u$ (lower), respectively.  }
    \label{fig:opt_n}
\end{figure}

Figure~\ref{fig:opt_n}, we separately evaluate how the optimal splitting-chunks quantity $n^*$ change with Alice's transmission power $P_{a}$, the message length $M$, and the square area size $u$ in the left three graphs. Meanwhile, we provide the corresponding catching probability $P_{ca}$ for each value of $n^*$. In each graph, we also evaluate the performance of Algorithm~\ref{Alg:2} by comparing it with the simulation.
The upper two graphs show that the splitting-chunks quantity reaches $15$ when $P_{a}$ is small, and remains at about $5$ when $P_{a}$ is large. Meanwhile, the corresponding overall catching probability slowly increases with $P_{a}$. 
The phenomenon can also be explained according to the observation in Figure~\ref{fig:change_L}. The middle two graphs show that $n^*$ and  $P_{ca}$ generally increase with $M$. Similar to the observation in Figure~\ref{fig:change_n}, splitting a message can bring more benefits from a large message. The lower graph shows that $n^*$ almost remains the same as the increase of $u$, and $P_{ca}$ decreases with the increasing $u$. It verifies the observation in Figure~\ref{fig:change_L} that $n^*$ is not sensitive to $u$.

\subsection{Cases Study}
Finally, we simulate two cases proposed in Section~\ref{subsec:case}. We assume that the total transmission slots for Alice is $100$, the new message arrives following the Poisson process with the rate $\lambda$ per minute, and the interval time between two transmission slots is $t_c=10$ minutes. 

\begin{figure}[htb]
    \centering
    \includegraphics[width=0.5\textwidth]{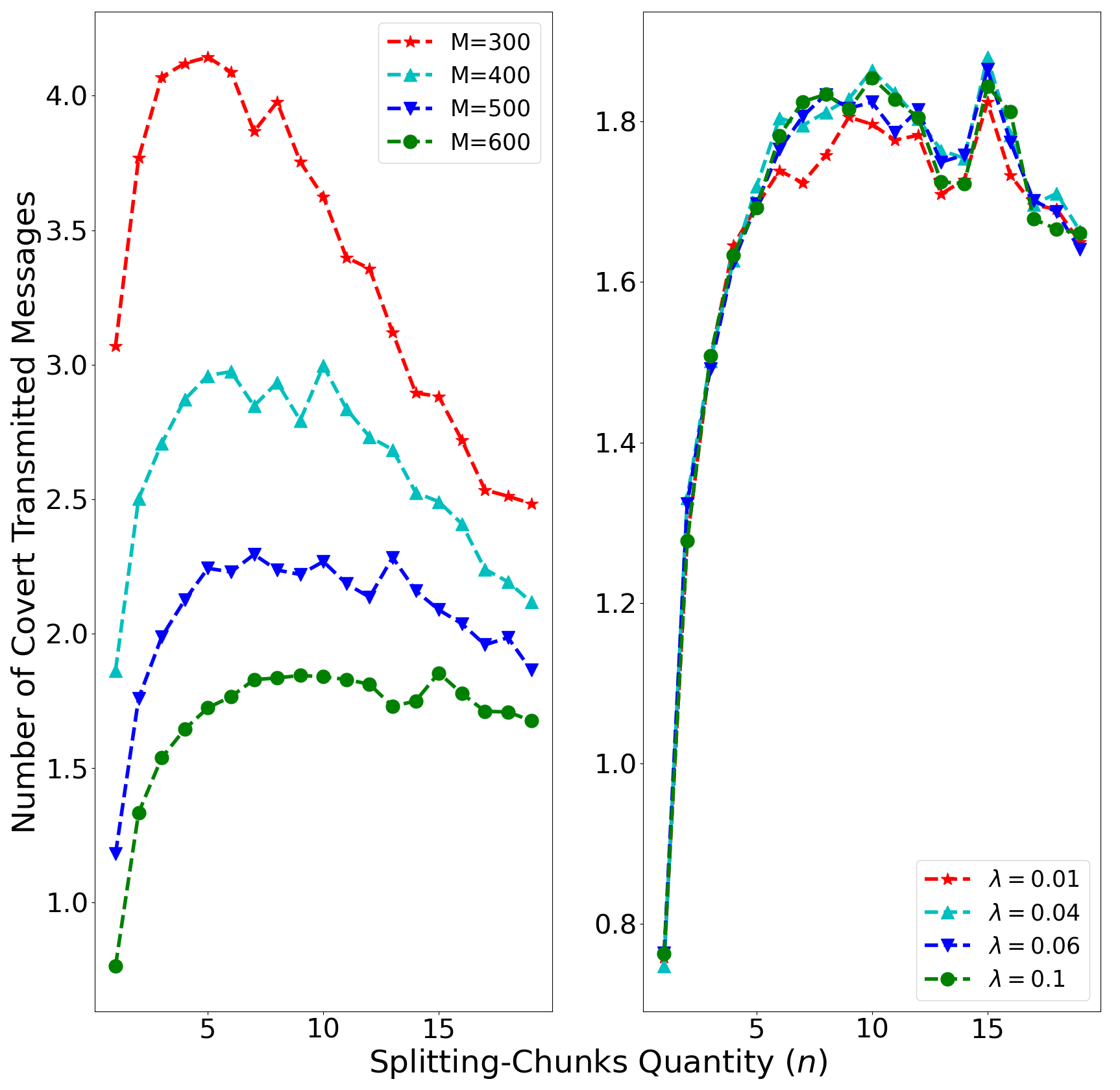}
    \caption{\textbf{Case 1:} The number of covert transmitted messages versus the splitting-chunks quantity $n$ for the message length $M$ (left) and the message arrival rate $\lambda$ (right), respectively. Willie's projected catching radius is set to $r_w^{\prime}=120$.}
    \label{fig:case1}
\end{figure}

Figure~\ref{fig:case1} shows the simulation results for Case 1.
We evaluate the number of covert transmitted messages versus the splitting-chunks quantity $n$ for different system parameters.  In the left graph, we consider the message length $M$ for four different values in $\{300, 400, 500, 600\}$.  The right graph considers the message arrival rate $\lambda$ for four different values in $\{0.1, 0.04, 0.06, 0.1\}$. Since Willie has a significant advantage in Case 1, we weaken Willie's catching radius to $r_w=120$ and keep the other parameters the same.
Both graphs show that the number of covert transmitted messages increases with $n$ and decreases thereafter. Splitting the message into smaller chunks works well for Case 1. Because Alice can not be caught by Willie, even once, the rest of the transmissions are not covert. Therefore, the better choice for Alice is to minimize the overall catching probability by increasing the number of chunks. We also find two observations to support the conclusion. In the left graph, the optimal $n$ increases with $M$, similar to the observation in the middle graph of Figure~\ref{fig:opt_n}. The right graph shows that the curves with different values of $\lambda$ almost perform the same, indicating that each message's transmission time is not the primary consideration in Case 1.

\begin{figure}[htb]
    \centering
    \includegraphics[width=0.5\textwidth]{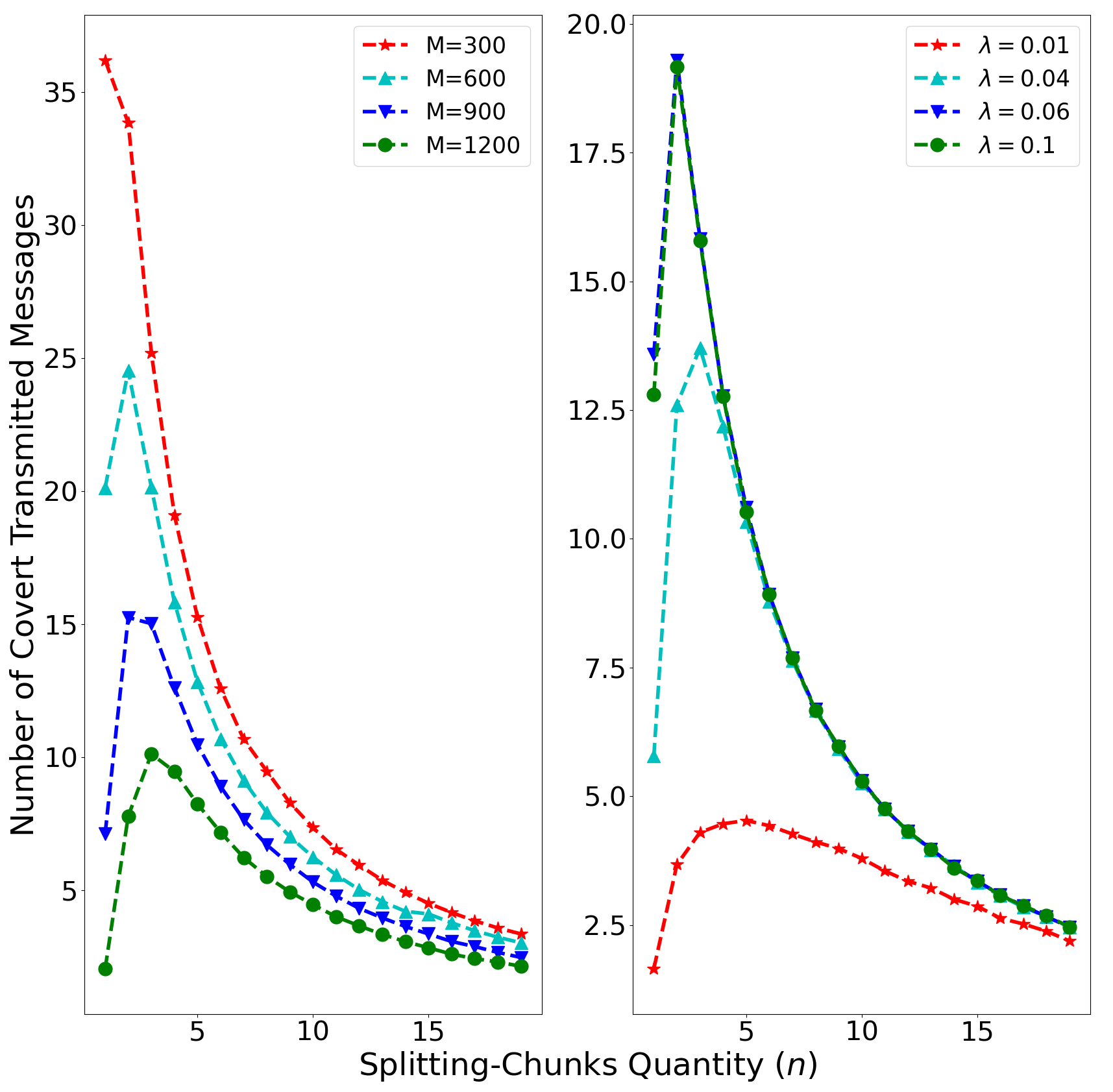}
    \caption{\textbf{Case 2:} The number of covert transmitted messages versus the splitting-chunks quantity $n$ for the message length $M$ (left) and the message arrival rate $\lambda$ (right), respectively. The message length is set to $M=900$.}
    \label{fig:case2}
\end{figure}

Figure~\ref{fig:case2} shows the simulation results for Case 2. We evaluate the number of covert transmitted messages versus the splitting-chunks quantity $n$ for different system parameters.  In the left graph, we consider the message length $M$ for four different values in $\{300, 600, 900, 1200\}$.  In the right graph, we consider the message arrival rate $\lambda$ for four different values in $\{0.1, 0.04, 0.06, 0.1\}$.
Both graphs show that the number of covert transmitted messages reaches the maximum with a small $n$. Although the splitting method is still working for Case 2, optimizing the transmission time for each message is also essential. Because Alice can still transmit the message covertly after being caught by Willie once. Therefore, transmitting more messages within the constraint time may become a better choice.

\section{Conclusions}
\label{sec:conclusions}
This paper considers a novel covert communication scenario in which Alice is the ground user intending to transmit the unauthorized message covertly to Bob, the LEO satellite; Willie is a UAV patrolling in the sky to detect the unauthorized message in a square area. Unlike the classic covert communication model, we consider that Alice's location does not limit satellite communication and assume that Alice and Willie realize each other's existence but do not know each other's location. We define the transmission as covert if Willie fails to detect or if he can not locate Alice during the transmission. In this scenario, we explored how the detection window size affects the catching probability from Willie's perspective and proposed an algorithm to approximate the optimal detection window size. Then, we also analyze how many chunks should be split to minimize the overall catching probability from Alice's perspective and propose an algorithm to approximate the number of chunks. The simulation results show that the detection window size and the splitting-chunks quantity significantly affect the catching and overall catching probabilities. 

\bibliographystyle{IEEEtran}
\bibliography{covert}

\begin{thebibliography}{10}
\providecommand{\url}[1]{#1}
\csname url@samestyle\endcsname
\providecommand{\newblock}{\relax}
\providecommand{\bibinfo}[2]{#2}
\providecommand{\BIBentrySTDinterwordspacing}{\spaceskip=0pt\relax}
\providecommand{\BIBentryALTinterwordstretchfactor}{4}
\providecommand{\BIBentryALTinterwordspacing}{\spaceskip=\fontdimen2\font plus
\BIBentryALTinterwordstretchfactor\fontdimen3\font minus \fontdimen4\font\relax}
\providecommand{\BIBforeignlanguage}[2]{{%
\expandafter\ifx\csname l@#1\endcsname\relax
\typeout{** WARNING: IEEEtran.bst: No hyphenation pattern has been}%
\typeout{** loaded for the language `#1'. Using the pattern for}%
\typeout{** the default language instead.}%
\else
\language=\csname l@#1\endcsname
\fi
#2}}
\providecommand{\BIBdecl}{\relax}
\BIBdecl

\bibitem{alsabah20216g}
M.~Alsabah, M.~A. Naser, B.~M. Mahmmod, S.~H. Abdulhussain, M.~R. Eissa, A.~Al-Baidhani, N.~K. Noordin, S.~M. Sait, K.~A. Al-Utaibi, and F.~Hashim, ``6g wireless communications networks: A comprehensive survey,'' \emph{IEEE Access}, vol.~9, pp. 148\,191--148\,243, Nov. 2021.

\bibitem{he2017covert}
B.~He, S.~Yan, X.~Zhou, and V.~K. Lau, ``On covert communication with noise uncertainty,'' \emph{IEEE Communications Letters}, vol.~21, no.~4, pp. 941--944, Apr. 2017.

\bibitem{chen2023covert}
X.~Chen, J.~An, Z.~Xiong, C.~Xing, N.~Zhao, F.~R. Yu, and A.~Nallanathan, ``Covert communications: A comprehensive survey,'' \emph{IEEE Communications Surveys \& Tutorials}, vol.~25, no.~2, pp. 1173--1198, 2nd Quart. 2023.

\bibitem{peng2022covert}
P.~Peng and E.~Soljanin, ``Covert, low-delay, coded message passing in mobile (iot) networks,'' \emph{IEEE Transactions on Information Forensics and Security}, vol.~17, pp. 599--611, Feb. 2022.

\bibitem{bash2013limits}
B.~A. Bash, D.~Goeckel, and D.~Towsley, ``Limits of reliable communication with low probability of detection on awgn channels,'' \emph{IEEE journal on selected areas in communications}, vol.~31, no.~9, pp. 1921--1930, Sep. 2013.

\bibitem{bash2015hiding}
B.~A. Bash, D.~Goeckel, D.~Towsley, and S.~Guha, ``Hiding information in noise: Fundamental limits of covert wireless communication,'' \emph{IEEE Communications Magazine}, vol.~53, no.~12, pp. 26--31, Dec. 2015.

\bibitem{du2022performance}
H.~Du, D.~Niyato, Y.-A. Xie, Y.~Cheng, J.~Kang, and D.~I. Kim, ``Performance analysis and optimization for jammer-aided multiantenna uav covert communication,'' \emph{IEEE Journal on Selected Areas in Communications}, vol.~40, no.~10, pp. 2962--2979, Otc. 2022.

\bibitem{he2023channel}
R.~He, J.~Chen, G.~Li, H.~Wang, Y.~Xu, W.~Yang, Y.~Jiao, and W.~He, ``Channel-aware jammer selection and power control in covert communication,'' \emph{IEEE Transactions on Vehicular Technology}, vol.~73, no.~2, pp. 2266--2279, Feb. 2023.

\bibitem{chen2024achieving}
X.~Chen, F.~Gao, M.~Qiu, J.~Zhang, F.~Shu, and S.~Yan, ``Achieving covert communication with a probabilistic jamming strategy,'' \emph{IEEE Transactions on Information Forensics and Security}, vol.~19, pp. 5561 -- 5574, May 2024.

\bibitem{yu2024covert}
H.~Yu, J.~Yu, J.~Liu, Y.~Li, N.~Ye, K.~Yang, and J.~An, ``Covert satellite communication over overt channel: A randomized gaussian signalling approach,'' \emph{IEEE Transactions on Aerospace and Electronic Systems}, Early Access, Oct. 2024, \uppercase{DOI}: 10.1109/TAES.2024.3475994.

\bibitem{wang2024star}
Q.~Wang, S.~Guo, C.~Wu, C.~Xing, N.~Zhao, D.~Niyato, and G.~K. Karagiannidis, ``Star-ris aided covert communication in uav air-ground networks,'' \emph{IEEE Journal on Selected Areas in Communications}, vol.~43, no.~1, pp. 245 -- 259, Jan. 2025.

\bibitem{li2024covert}
X.~Li, Z.~Tian, W.~He, G.~Chen, M.~C. Gursoy, S.~Mumtaz, and A.~Nallanathan, ``Covert communication of star-ris aided noma networks,'' \emph{IEEE Transactions on Vehicular Technology}, vol.~73, no.~6, pp. 9055--9060, Jun. 2024.

\bibitem{feng2023radio}
W.~Feng, Y.~Lin, Y.~Wang, J.~Wang, Y.~Chen, N.~Ge, S.~Jin, and H.~Zhu, ``Radio map-based cognitive satellite-uav networks towards 6g on-demand coverage,'' \emph{IEEE Transactions on Cognitive Communications and Networking}, vol.~10, no.~3, pp. 1075--1089, Jun. 2024.

\bibitem{you2024ubiquitous}
L.~You, Y.~Zhu, X.~Qiang, C.~G. Tsinos, W.~Wang, Z.~Gao, and B.~Ottersten, ``Ubiquitous integrated sensing and communications for massive mimo leo satellite systems,'' \emph{IEEE Internet of Things Magazine}, vol.~7, no.~4, pp. 30--35, Jul. 2024.

\bibitem{wang2024sustainable}
F.~Wang, S.~Zhang, J.~Shi, Z.~Li, and T.~Q. Quek, ``Sustainable uav mobility support in integrated terrestrial and non-terrestrial networks,'' \emph{IEEE Transactions on Wireless Communications}, vol.~23, no.~11, pp. 17\,115 -- 17\,128, Nov. 2024.

\bibitem{capez2024use}
G.~M. Capez, M.~A. C{\'a}ceres, R.~Armellin, C.~P. Bridges, J.~A. Fraire, S.~Frey, and R.~Garello, ``On the use of mega constellation services in space: Integrating leo platforms into 6g non-terrestrial networks,'' \emph{IEEE Journal on Selected Areas in Communications}, vol.~42, no.~12, pp. 3490--3504, Dec. 2025.

\bibitem{wang2023satellite}
S.~Wang and Q.~Li, ``Satellite computing: Vision and challenges,'' \emph{IEEE Internet of Things Journal}, vol.~10, no.~24, pp. 22\,514--22\,529, Dec. 2023.

\bibitem{ma2021uav}
T.~Ma, H.~Zhou, B.~Qian, N.~Cheng, X.~Shen, X.~Chen, and B.~Bai, ``Uav-leo integrated backbone: A ubiquitous data collection approach for b5g internet of remote things networks,'' \emph{IEEE Journal on Selected Areas in Communications}, vol.~39, no.~11, pp. 3491--3505, Nov. 2021.

\bibitem{peng2024blocked}
P.~Peng, T.~Xu, X.~Chen, C.~C. Zarakovitis, and C.~Wu, ``Blocked job offloading based computing resources sharing in leo satellite networks,'' \emph{IEEE Internet of Things Journal}, vol.~12, no.~2, pp. 2287--2290, Jan. 2025.

\bibitem{he2024direct}
Y.~He, Y.~Xiao, S.~Zhang, M.~Jia, and Z.~Li, ``Direct-to-smartphone for 6g ntn: Technical routes, challenges, and key technologies,'' \emph{IEEE Network}, vol.~38, no.~4, pp. 128 -- 135, Jul. 2024.

\bibitem{feng2024covert}
S.~Feng, X.~Lu, S.~Sun, E.~Hossain, G.~Wei, and Z.~Ni, ``Covert communication in large-scale multi-tier leo satellite networks,'' \emph{IEEE Transactions on Mobile Computing}, vol.~23, no.~12, pp. 11\,576--11\,587, May 2024.

\bibitem{jia2025robust}
H.~Jia, Y.~Wang, W.~Wu, and J.~Yuan, ``Robust transmission design for covert satellite communication systems with dual-csi uncertainty,'' \emph{IEEE Internet of Things Journal}, Early Access, Mar. 2025, \uppercase{DOI}: 10.1109/JIOT.2025.3550098.

\bibitem{chen2021uav}
X.~Chen, M.~Sheng, N.~Zhao, W.~Xu, and D.~Niyato, ``Uav-relayed covert communication towards a flying warden,'' \emph{IEEE Transactions on Communications}, vol.~69, no.~11, pp. 7659--7672, Nov. 2021.

\bibitem{zhou2019joint}
X.~Zhou, S.~Yan, J.~Hu, J.~Sun, J.~Li, and F.~Shu, ``Joint optimization of a uav's trajectory and transmit power for covert communications,'' \emph{IEEE Transactions on Signal Processing}, vol.~67, no.~16, pp. 4276--4290, Jul. 2019.

\bibitem{song2023ris}
D.~Song, Z.~Yang, G.~Pan, S.~Wang, and J.~An, ``Ris-assisted covert transmission in satellite--terrestrial communication systems,'' \emph{IEEE Internet of Things Journal}, vol.~10, no.~22, pp. 19\,415--19\,426, Feb. 2023.

\bibitem{liu2024covert}
P.~Liu, J.~Si, Z.~Li, and N.~Al-Dhahirl, ``Covert communications for cognitive satellite terrestrial networks,'' in \emph{2024 IEEE Wireless Communications and Networking Conference (WCNC)}.\hskip 1em plus 0.5em minus 0.4em\relax IEEE, Jul. 2024, pp. 1--6.

\bibitem{wang2022covert}
C.~Wang, X.~Chen, J.~An, Z.~Xiong, C.~Xing, N.~Zhao, and D.~Niyato, ``Covert communication assisted by uav-irs,'' \emph{IEEE Transactions on Communications}, vol.~71, no.~1, pp. 357--369, Jan. 2022.

\bibitem{jiao2024uav}
L.~Jiao, X.~Chen, L.~Xu, N.~Deng, N.~Zhao, and X.~Wang, ``Uav-relayed finite-blocklength covert communication with channel estimation,'' \emph{IEEE Transactions on Vehicular Technology}, vol.~73, no.~6, pp. 9032--9037, Jun. 2024.

\bibitem{zhou2021three}
X.~Zhou, S.~Yan, D.~W.~K. Ng, and R.~Schober, ``Three-dimensional placement and transmit power design for uav covert communications,'' \emph{IEEE Transactions on Vehicular Technology}, vol.~70, no.~12, pp. 13\,424--13\,429, Dec. 2021.

\bibitem{zheng2019multi}
T.-X. Zheng, H.-M. Wang, D.~W.~K. Ng, and J.~Yuan, ``Multi-antenna covert communications in random wireless networks,'' \emph{IEEE Transactions on Wireless Communications}, vol.~18, no.~3, pp. 1974--1987, Mar. 2019.

\bibitem{jiang2021resource}
X.~Jiang, Z.~Yang, N.~Zhao, Y.~Chen, Z.~Ding, and X.~Wang, ``Resource allocation and trajectory optimization for uav-enabled multi-user covert communications,'' \emph{IEEE Transactions on Vehicular Technology}, vol.~70, no.~2, pp. 1989--1994, Feb. 2021.

\bibitem{jiang2021covert}
X.~Jiang, X.~Chen, J.~Tang, N.~Zhao, X.~Y. Zhang, D.~Niyato, and K.-K. Wong, ``Covert communication in uav-assisted air-ground networks,'' \emph{IEEE Wireless Communications}, vol.~28, no.~4, pp. 190--197, Aug. 2021.

\bibitem{deng2024joint}
D.~Deng, W.~Zhou, X.~Li, D.~B. Da~Costa, D.~W.~K. Ng, and A.~Nallanathan, ``Joint beamforming and uav trajectory optimization for covert communications in isac networks,'' \emph{IEEE Transactions on Wireless Communications}, vol.~24, no.~2, pp. 1016--1030, Feb. 2025.

\bibitem{weisstein2004square}
E.~W. Weisstein, ``Square line picking,'' \emph{https://mathworld. wolfram. com/}, 2004.

\end{thebibliography}

\end{document}